\documentclass[aps,pre,twocolumn]{revtex4-2}
\usepackage{amsfonts,amssymb,amsmath,bm}
\usepackage{graphicx}
\usepackage{hyperref}
\usepackage{color}
\usepackage{physics}
\usepackage[dvipsnames]{xcolor}

\begin{document}

\title{Effect of active loop extrusion on the two-contact correlations in the interphase chromosome}
\date\today

\author{Dmitry Starkov$^{1,2}$ and Sergey Belan$^{1,3}$}
\email{sergb27@yandex.ru}
\affiliation{$^{1}$Landau Institute for Theoretical Physics, Russian Academy of Sciences, 1-A Akademika Semenova av., 142432 Chernogolovka, Russia}
\affiliation{$^{2}$National Research University Higher School of Economics, Faculty of Mathematics, Usacheva 6, 119048 Moscow, Russia}
\affiliation{$^{3}$National Research University Higher School of Economics, Faculty of Physics, Myasnitskaya 20, 101000 Moscow, Russia}

\begin{abstract} 
The population-averaged contact maps generated by the chromosome conformation capture technique provide important information about the average frequency of contact between pairs of chromatin loci as a function of the genetic distance between them. 
However, these datasets do not tell us anything about the joint statistics of simultaneous contacts between genomic loci in individual cells. 
This kind of statistical information can be extracted using the single-cell Hi-C method, which is capable of detecting a  large fraction of simultaneous contacts within a single cell, as well as through modern methods of fluorescent labeling and super-resolution imaging. 
Motivated by the prospect of the imminent availability of relevant experimental data, in this work we theoretically model the joint statistics of pairs of contacts located along a line perpendicular to the main diagonal of the single-cell contact map.
The analysis is performed within the framework of an ideal polymer model with quenched disorder of random loops, which, as previous studies have shown, allows to take into account the influence of the loop extrusion process on the conformational properties of interphase chromatin.
\end{abstract}

\maketitle

\section{Introduction}

A series of recent experiments \cite{ganji2018real,golfier2020cohesin,kong2020human,davidson2019dna,kim2019human,ryu2020condensin,banigan2020loop} have shown that the structural maintenance of chromosomes (SMC) proteins, such as condensin and cohesin, when binding to DNA can exhibit (ATP-dependent) motor activity leading to progressive growth of DNA loops.
Due to these works, the loop extrusion process, originally proposed as a hypothetical molecular mechanism, has  received direct experimental confirmation \cite{kimura199913s,nasmyth2001disseminating,riggs1990dna}.

\begin{figure}[t]
    \centering
    \includegraphics[scale=0.38]{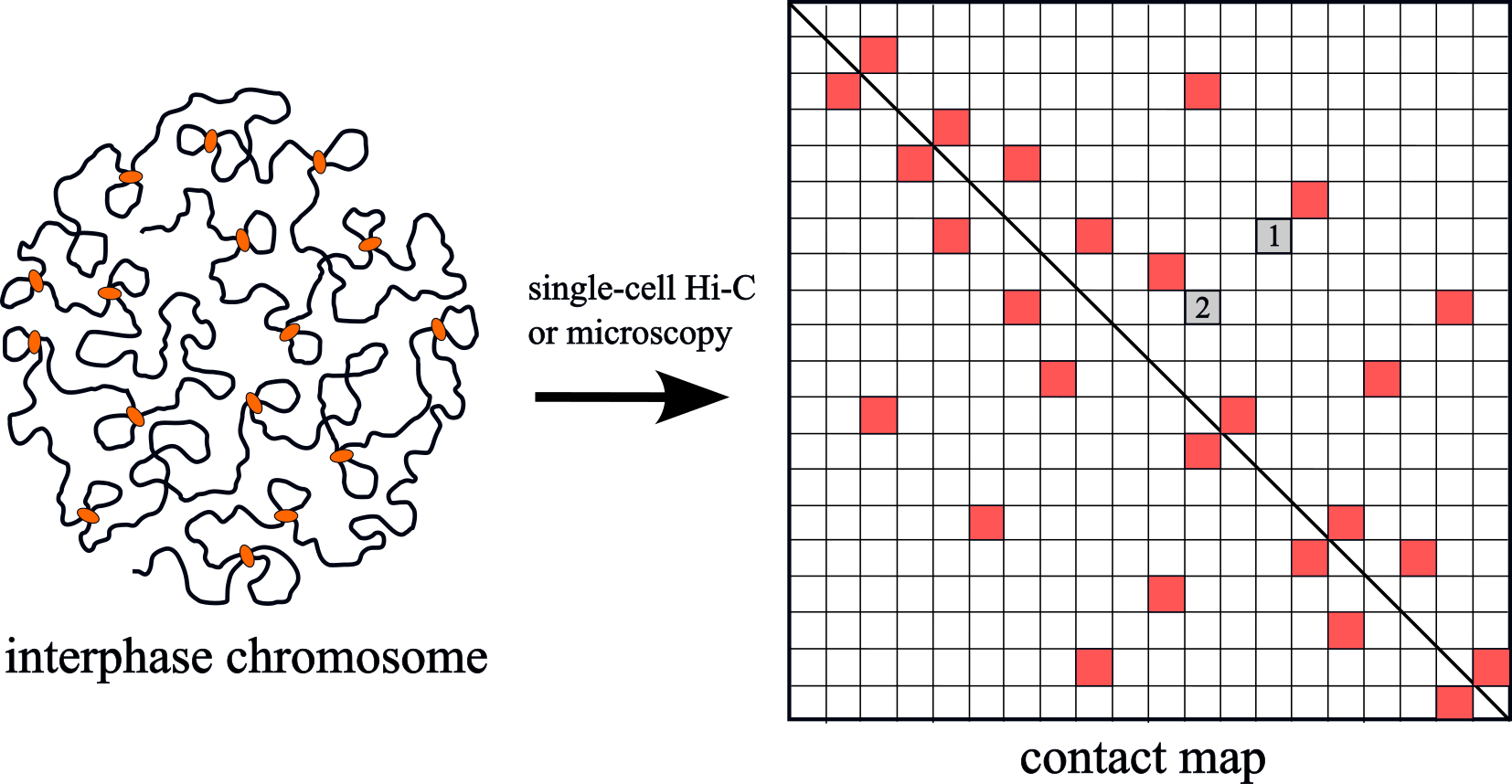}
    \caption{Left: Interphase chromosome of higher eukaryotes can be modeled as polymer chain with an array of sufficiently sparse random loops which are in practice produced due to motor activity of cohesin proteins. Right: Modern experimental techniques 
    allow to construct single-cell contact map depicting the pattern of simultaneous contacts - both random and cohesin-induced. The size of the pixels is determined by the resolution of experimental technique and can be also controlled at the stage of data analysis via coarse-graining. We are interested in joint statistics of contacts belonging to the line perpendicular to the main diagonal of the contact map, see e.g. pixels 1 and 2.}
    \label{Fig: 1}
\end{figure}

The influence of loop extrusion machinery on the conformational properties of chromatin can be modeled within the framework of ideal  chain model with array of random loops, see Fig. \ref{Fig: 1}. 
In \cite{polovnikov2023crumpled} we showed that this simple model qualitatively correctly reproduces the specific experimental profile of pairwise contact frequency in its dependence on the genomic distances at the scales up to $1Mb$.
In addition, we have previously used this model to extract theoretical predictions regarding the conditional contact probability \cite{belan2022influence} and the mean square of physical distance \cite{belan2023footprints} between pairs of genomic loci.
Generalization of the model to the more realistic case of a non-ideal polymer was discussed in \cite{polovnikov2023crumpled} and \cite{slavov2023intrachain}.

The analysis presented in this theoretical work is aimed at further development of this minimalistic model with the aim of describing a larger body of potentially accessible experimental data.
More specifically, we describe the joint statistics of the two contacts lying on the line perpendicular to the main diagonal of the contact map.
The  opportunity to extract this kind of data for chromatin in a living cell appeared thanks to the development of the single-cell Hi-C method \cite{cattoni2017single,gassler2017mechanism,nagano2013single,stevens20173d,nagano2015single,ramani2020sci,galitsyna2021single,kos2021perspectives,zhang2022multiscale}.
Unlike the population averaged contact maps, generated by  Hi-C method which detect only a small fraction of all contacts in single cells, this technique provides high coverage of contacts at the single-cell level (see Fig. \ref{Fig: 1}).
Besides, the joint statistics of two contacts is potentially available due to modern tools for high-throughput super-resolution imaging enabling direct visualization of the spatial positions of many genomic loci in individual cells \cite{cattoni2017single,
ou2017chromemt,bintu2018super,nir2018walking,boettiger2020advances,kempfer2020methods, su2020genome,liu2020multiplexed, xie2021single,li2021nanoscale,gabriele2022dynamics}.


\section{Model formulation}
\label{sec: model}

The interphase chromatin can be modelled as a polymer chain with dynamic array of cohesin-mediated loops, see Fig. \ref{Fig: 1}.
The growth of a loop starts when a cohesin complex binds to chromatin, and stops either at the moment of dissociation, when free cohesin returns to the nucleoplasm, or as a result of blocking of motor activity by another cohesin complex or by CTCF protein \cite{fudenberg2017emerging}.
The mean processivity $\lambda$ of extrusion is defined as the average contour length of the loop generated by a cohesin in the absence of interaction with any unit that can block the loop growth.
Another important characteristic of extrusion process is the average contour distance $d$ between the binding sites of neighboring cohesin complexes.
For interphase conditions, the inequality $\lambda/d\lesssim 1$ is valid \cite{polovnikov2023crumpled,Goloborodko_2016}, so that for a typical chain conformation most of the cohesin-mediated loops are separated by fairly large gaps, as shown in Fig. \ref{Fig: 1}, whereas
 the fraction of nested and spliced loops is relatively small. 
Due to the stochasticity of the binding and unbinding events associated with exchange by cohesin between chromatin and nucleoplasm  the resulting contour  lengths  of loops should be considered a random variables.
This argument together with irregular pattern of the binding sites of SMC-proteins mean that the contour lengths of the chain segments enclosed between adjacent loops should also be treated statistically.
In our analytical calculations we will consider the asymptotic limit
$\lambda/d\ll 1$.
In this case, the contour lengths of loops, $L$, and gaps, $h$, can be considered as independent random variables with exponential probability density functions
\begin{equation}
\label{loops_pdf}
    p_{\text {loop }}(L)=\frac{1}{\lambda} \exp \left(-\frac{L}{\lambda}\right),
\end{equation}
and 
\begin{equation}
\label{gaps_pdf}
    p_{\text {gap }}(h)=\frac{1}{d} \exp \left(-\frac{h}{d}\right).
\end{equation}

Next, we will assume that chromatin is an ideal chain with the Kuhn segment length $l_{\text{eff}}$ \cite{GKh_1994}. 
The conformation of the chain changes randomly over time due to thermal fluctuations and the motor activity of loop-extruding complexes.
Focusing on the slow extrusion limit, we will assume that the loop disorder is frozen, so that random conformations of the polymer chain belong to equilibrium ensemble.
More specifically, this means that chain conformation can be thought of as alternating free Brownian paths and Brownian bridges \cite{belan2023footprints}.
 


\begin{table}[t!]
\begin{center}
\begin{tabular}{|l|c|c|c|}
\hline
Event & \multicolumn{1}{l|}{Probability}  \\ \hline
$\sigma_{i_1j_1}=1$, $\sigma_{i_2j_2}=1$      & $C(s_1,s_2)$                                                     \\ \hline
$\sigma_{i_1j_1}=1$, $\sigma_{i_2j_2}=0$      & $P(s_1)-C(s_1,s_2)$                                                    \\ \hline
$\sigma_{i_1j_1}=0$, $\sigma_{i_2j_2}=1$      & $P(s_2)-C(s_1,s_2)$                                                   \\ \hline
$\sigma_{i_1j_1}=0$, $\sigma_{i_2j_2}=0$      & $1-P(s_1)-P(s_2)+C(s_1,s_2)$                                                 \\ \hline
\end{tabular}
\end{center}
\caption{The probabilities of different values of a pair of random variables $\sigma_{i_1j_1}\sigma_{i_2j_2}$ can be expressed via the pairwise contact probability $P$ and the pair correlation function $C$.} 
\end{table}


The probability density function of a random vector $\vec R$ connecting two points of an equilibrium ideal chain belonging to the gap region between the cohesin-mediated loop is determined by the expression
(see, e.g., \cite{GKh_1994})
\begin{equation}
\label{pfree}
P_{\text {free }}(\vec{R}\mid s)=\frac{1}{\left(2 \pi \sigma_{\text {free }}^2(s)\right)^{3 / 2}} \exp \left(-\frac{R^2}{2 \sigma_{\text {free }}^2(s)}\right),
\end{equation}
where $\sigma_{\text{free}}^2(s)=\frac{l_{\text{eff}}^2s}{3}$ and $s\gg l_{\text{eff}}$ denotes the contour distance between points of interest. 

For the probability distribution of the random vector $\vec R$ connecting two points belonging to a cohesin-mediated loop one has (see, e.g., \cite{GKh_1994})
\begin{equation}
\label{pcoil}
P_{\mathrm{coil}}(\vec{R} \mid s, L)=\frac{1}{\left(2 \pi \sigma_{\text {coil }}^2(s, L)\right)^{3 / 2}} \exp \left(-\frac{R^2}{2 \sigma_{\text {coil }}^2(s, L)}\right).
\end{equation}
Here $\sigma_{\text {coil }}^2(s, L)=\frac{l_{\text{eff}}^2s(L-s)}{3L}$, $L$  is the contour length of the loop,  and $s$ is the  contour distance between points of interest (assuming $l_{\text{eff}} s < L$).

Formulas (\ref{loops_pdf}), (\ref{gaps_pdf}), (\ref{pfree}) and (\ref{pcoil}) are sufficient to assign the probability measure to various conformations of ideal chain with array of random loops.


\section{Metrics of interest and sketch of calculations}
\label{sec: calculations}

Let us introduce the indicator of contact, i.e.  
a binary random variable $\sigma_{ij}$, which is equal to unity if there is contact between points of the polymer chain with contour coordinates $i$ and $j$, and equal to zero otherwise.
This can be formally expressed as
\begin{equation}
\label{contact_indicator}
\sigma_{ij}=I(|\vec r_{i}-\vec r_{j}|\le a),
\end{equation}
where the vectors $\vec r_{i}$ and $\vec r_{j}$ determine the positions of the loci $i$ and $j$ in space,  $I(\dots)$ is an indicator function,  variable, and  $a$ denotes the contact radius whose value depends on the details of the experimental technique. 

It follows from definition (\ref{contact_indicator}) that probability $P(s)$ to find two points separated by genomic distance $s=|i-j|$    in contact is given by 
the expected value of the contact indicator, i.e.
\begin{eqnarray}
P(s)=\langle\sigma_{ij}\rangle. 
\end{eqnarray}
Next, the pair correlation function 
of contact indicators
\begin{equation}
C(s_1,s_2)=\langle \sigma_{i_1j_1}\sigma_{i_2j_2}\rangle,
\end{equation}
gives the probability 
to find a contact between both pairs of points: $(i_1,j_1)$ and $(i_2,j_2)$.
As shown in Table 1, two functions $P(s)$ and $C(s_1,s_2)$ completely characterise  
joint statistics of any two pixels in single-cell contact map.
In this paper we calculate these objects analytically using  the  ideal chain model with quenched disorder of loops described in the previous section.
For the sake of simplicity of the calculations we focus on the scenarios $i_1\le i_2\le j_2\le j_1$ and additionally assume the condition $i_2-i_1=j_1-j_2$.
In other words, our analysis considers pairs of contacts belonging to the line perpendicular to the main diagonal of the contact map, see Fig. \ref{Fig: 1}.

Let us briefly describe the  key steps of calculations required to arrive at $P(s)$ and $C(s_1,s_2)$.

\begin{figure}[t!]
    \centering
    \includegraphics[scale=0.15]{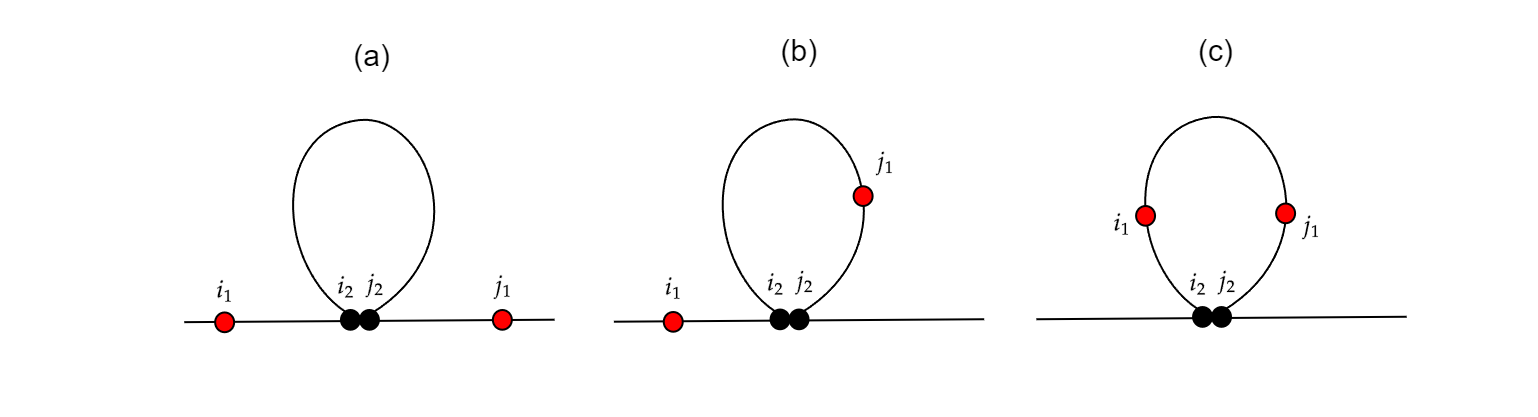}
    \caption{Three types of one-loop diagrams contributing to contact probability  $P(s)$ between  two points of  ideal chain with  disorder of rare random loops.
    The bases of the cohesin-mediated loops are depicted as black dots.}
    \label{Fig: 2}
\end{figure}

\begin{figure}[t!]
    \centering
    \includegraphics[scale=0.16]{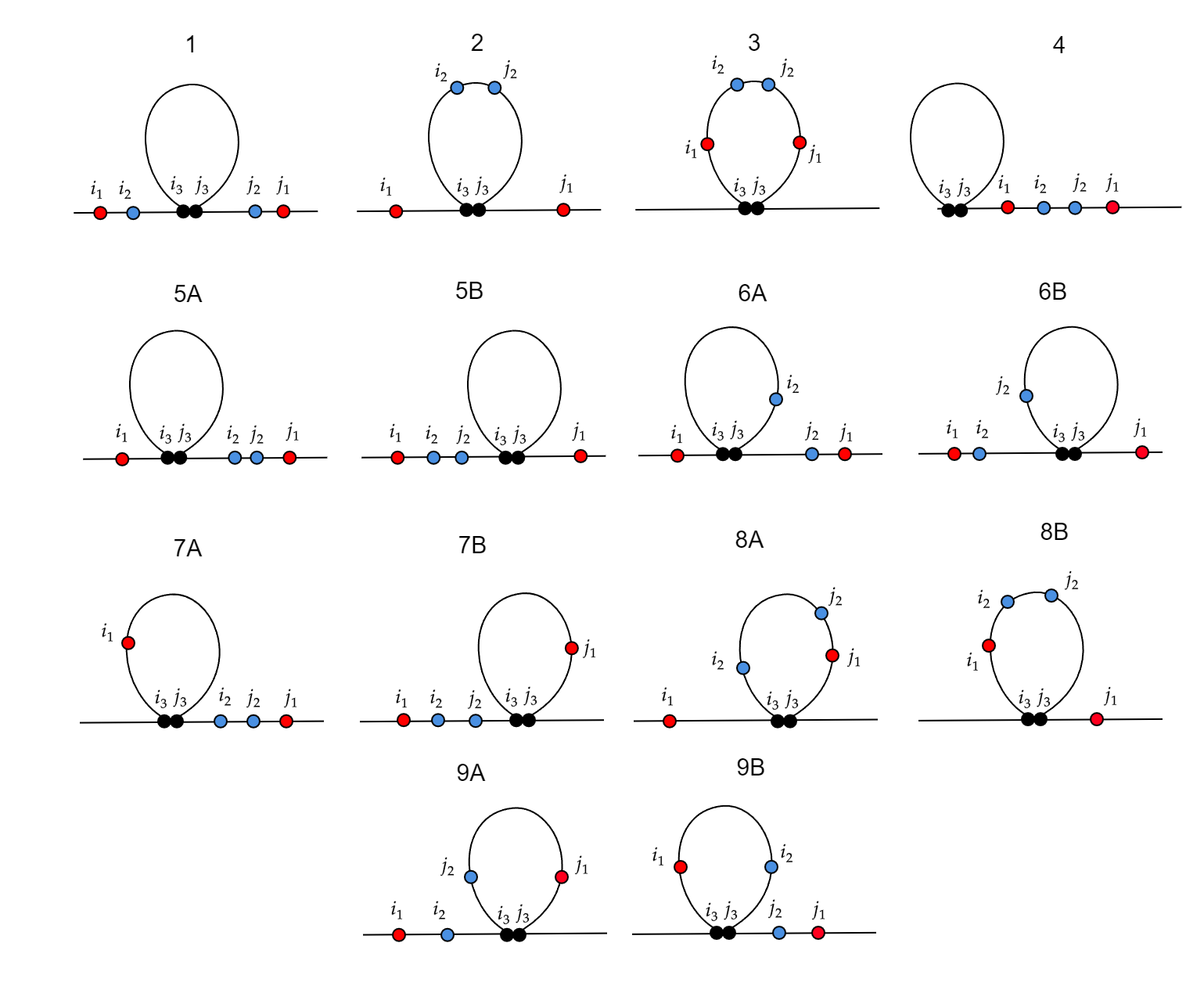}
    \caption{Fourteen classes of one-loop diagrams contributing to pair correlation function $C(s_1,s_2)$ of the ideal chain with  disorder of rare random loops.
    The bases of the cohesin-mediated loops are depicted as black dots.}
    \label{Fig: 3}
\end{figure}

Strictly speaking, there is an infinitely large set of scenarios for the mutual arrangement of points $i$, $j$ and the bases of loops formed through the
motor activity of SMC complexes.
The route to the integral representation for contact probability accounting all types of loop patterns at $\lambda/d\lesssim 1$ was described in  Ref. \cite{polovnikov2023crumpled}.
The resulting expression however contains multiple integrals which should be evaluated numerically.
In the rare loop limit adopted here the relatively simple analytical answer for $P(s)$  can be obtained.
Namely, if $\lambda/d\ll 1$, then in the region $s\ll 1$ the leading contribution to the expectation $\langle \sigma_{ij}\rangle$ is determined by 3 one-loop diagrams, see Fig. \ref{Fig: 2}, i.e.
\begin{eqnarray}
\label{P_decomposition}
P(s)=\sum_{\alpha=a,b,c}\langle  {\cal Q}_{\alpha}(s|\{ A\}_\alpha)\rangle_{\text{loops}},
\end{eqnarray}
where $\alpha$ enumerates the diagrams, ${\cal Q}_\alpha(s|\{ A\}_\alpha)$ - is the conditional expectation obtained by averaging of  $\sigma_{ij}$ over thermal noise at fixed pattern of random loops, $\{ A\}_\alpha$ represents the set of random variables parametrising the corresponding diagram, and $\langle\cdots\rangle_{\text{loops}}$
denotes the averaging over the variables $\{ A\}_\alpha$.
The conditional expectations associated with different diagrams are given by
\begin{eqnarray}
   &&{\cal Q}_\alpha(s|\{ A\}_\alpha)\approx\frac{4}{3} \pi a^{3}\int {\cal P}_\alpha(\vec R|\{A\}_\alpha;s)\delta(\vec R)d^3R=\\
   \label{Q_alpha}
   &&=\frac{4}{3}\pi a^{3}{\cal P}_\alpha(0|\{A\}_\alpha;s). 
\end{eqnarray}
Here ${\cal P}_{\alpha}(\vec R|\{A\}_\alpha;s)$ - is the probability density function of the vector   $\vec R=\vec r_{i}-\vec r_{j}$ conditioned to the particular parameters of cohesin-mediated loop.
The variable $s$ enters this function  as a parameter.
While deriving Eq. (\ref{Q_alpha}) we assumed  that $a\ll \sqrt{\langle R^2\rangle}$.


Next,  assuming again the rare loop limit $\lambda/d\ll 1$, we find that at scales $s_1,s_2\ll d$ the leading contribution to the pair correlator is determined by 14 one-loop diagrams, see Fig. \ref{Fig: 3}, i.e.
\begin{eqnarray}
\label{C_representation}
C(s_1,s_2)=\sum_{\beta=1}^{14}\langle {\cal C}_{\beta}(s_1,s_2|\{ B\}_\beta)\rangle_{\text{loops}},
\end{eqnarray}
where $\beta$ enumerates the diagrams, ${\cal C}_\beta(s_1,s_2|\{ B\}_\beta)$ - is the conditional pair correlation function obtained by averaging of the product $\sigma_{i_1j_1}\sigma_{i_2j_2}$ over thermal noise at the fixed pattern of random loops, $\{ B\}_\beta$ represents the set of random variables parametrising the corresponding diagram, and $\langle\cdots\rangle_{\text{loops}}$
denotes the averaging over the variables $\{ B\}_\beta$.
The contributions coming from different diagrams can be expressed as
\begin{widetext}
    \begin{eqnarray}
   &&{\cal C}_\beta(s_1,s_2|\{ A\}_\beta)\approx(\frac43\pi a^3)^2\int {\cal P}_\beta(\vec R_1,\vec R_2|\vec R_3=0, \{ B\}_\beta;s_1,s_2)\delta(\vec R_1)\delta(\vec R_2)d^3R_1d^3R_2=\\
   \label{C_alpha}
   &&=\frac{16}{9}\pi^2 a^6{\cal P}_\beta(\vec R_1=0,\vec R_2=0|\vec R_3=0,\{ B\}_\beta;s_1,s_2).
\end{eqnarray}
\end{widetext}
Here  ${\cal P}_\beta(\vec R_1,\vec R_2|\vec R_3=0,\{ B\}_\beta;s_1,s_2)$ denotes the joint probability density 
of vectors $\vec R_1\equiv \vec r_{i_1}-\vec r_{j_1}$ and $\vec R_2\equiv \vec r_{i_2}-\vec r_{j_2}$
conditioned to particular values of 
random parameters $\{ B\}_\beta$ and to the contact  $\vec R_3\equiv\vec j_3-\vec i_3=0$ between points $i_3$ and $j_3$ (see Fig. \ref{Fig: 3}).

Thus, our task reduces to finding conditional probabilities according to Eqs. (\ref{Q_alpha}) and (\ref{C_alpha}) and further averaging of the obtained expressions over the statistics of random loops. 
The averaging procedure implies integration of ${\cal Q}_\alpha$ (or ${\cal C}_\beta$) over set of variables $\{A \}_\alpha$ (or $\{B \}_\beta$) with some weights.
We implement this calculation program in Appendix.
The next section is devoted to presenting the final results.

Note that conditional averages ${\cal C}_\beta(s_1,s_2|\{ B\}_\beta)$ were the subject of calculations in the paper \cite{zhang2001probability}.
Here we calculate them by using a methodologically more simple approach. 
Comparison of our answers with those presented  in \cite{zhang2001probability} demonstrates the difference for diagrams 6A, 6B, 9A and 9B.
In contrast to \cite{zhang2001probability},  our expressions for these diagrams exhibit correct behaviour in limiting cases, which speaks in support of their correctness.





\section{Results and discussion}\label{sec: results}
The conditional probability density ${\cal P}_{\alpha}(\vec R|\{A\}_\alpha;s)$ is  given by Eqs. (\ref{app_P_alpha})-(\ref{s_eff_c_one}).
For the corresponding weights see Eqs. (\ref{Wa})-(\ref{Wc}).
Collecting all contributions according to decomposition (\ref{P_decomposition}) we find that pairwise contact probability is given by
  \begin{equation}
  \label{contact_prob_res}
P(s)= \frac{b}{s^{3/2}}\left(1+\frac{\lambda}{d}f \left(\frac{s}{\lambda} \right) \right),
\end{equation}
where
\begin{eqnarray}
&&f(s)= \left(3s+1\right)\left(2\sqrt{s}D\left(\sqrt{s}\right)-1\right)
+\\
&&+\sqrt{\pi}s^2e^{-s}\left(2U(\frac{1}{2};2;s)+U(\frac{1}{2};3;s)\right),
\end{eqnarray}
and $U(a;b;s)=\frac{1}{\Gamma(a)}\int_{0}^{+\infty} y^{a-1}(1+y)^{b-a-1}e^{-sy}dy$ -  confluent hypergeometric  function,
$D(s)=e^{-s^2}\int_{0}^{s}e^{y^2}dy$ -  Dawson integral.
We also introduced notation $b=\sqrt{\frac{6}{\pi}}\frac{a^3}{l_{\text{eff}}^{3/2}}$.
The intuition behind the structure of this result is quite transparent: at $\lambda/d\ll1$, the
leading contribution to contact probability $P(s)$ is determined
by the loops-free answer $\propto s^{-3/2}$, while the diagrams containing a single cohesin-mediated loop produce the linear-order correction in small parameter $\lambda/d$.

It is straightforward to show that   $f(s)\to0$ at $s\to 0$ so that the loop-free result $P_{0}(s)=bs^{-3/2}$  recovers at scales $s\ll\lambda$. 
This is because sufficiently short segments of ideal chain do not feel the loop constraints.
In the opposite limit $s\to \infty$ one obtains $f(s)\to 3/2$: the random loops reduce the effective contour distance between distant loci, thus, increasing their contact frequency. 
It should be immediately noted, however, that the one-loop approximation
underlying the above calculations is justified if $s\ll d$.
For this reason, one may expect that the large-scale behaviour dictated by Eq. (\ref{contact_prob_res}) is inaccurate. 
However, the direct calculations for the region $s\gg d$ give exactly the same
asymptotic $\lim_{s\to0} f(s) = 3/2$ as the one-loop approximation \cite{belan2023footprints}.

Next the conditional probability densities ${\cal P}_\beta(\vec R_1,\vec R_2|\vec R_3=0,\{ B\}_\beta;s_1,s_2)$ entering Eq. (\ref{C_alpha}) are determined by Eqs. (\ref{conditional:diag1}), (\ref{conditional:diag2}), (\ref{conditional:diag3}), (\ref{conditional:diag4}), (\ref{conditional:diag5}), (\ref{diagram5B}), (\ref{conditional:diag7}), (\ref{conditional:diag7b}),   (\ref{conditional:diag10a}) and (\ref{conditional:diag9})   in Appendix, whereas the statistical weights required to perform averaging over disorder of loops are specified by Eqs. (\ref{omega1})-(\ref{omega9}).
Summing up contributions coming from all diagrams depicted in Fig. \ref{Fig: 3} we obtain 
\begin{equation}
\label{eq:scaleless}
    C(s_1,s_2) = \frac{b^2}{(s_1 - s_2)^{3/2}  s_2^{3/2}} \left(1 + \frac{\lambda}{d}F\left(\frac{s_1}{\lambda},\frac{s_2}{\lambda} \right) \right),
\end{equation}
where $F(s_1,s_2)$ is given by Eq. (\ref{F12}) in Appendix.
We see again that one-loop diagrams generate the linear correction in small parameter $\lambda/d$ to loops-free answer.
As is clear from Eq. (\ref{eq:scaleless}) and the left panel of Fig. \ref{Fig: 5}, for a fixed value of $s_2$ correlations decay  as a power law with an exponent $-3/2$ at large $s_1$. 
Being considered as a function of $s_2$,  correlator $C(s_1,s_2)$ is non-monotonic and attains its minimum at $s_2\approx s_1/2$, see left panel of Fig. \ref{Fig: 6}. 
This is explained by the structure of the loops-free correlation function $C_{0}(s_1,s_2)=\frac{b^2}{(s_1 - s_2)^{3/2}  s_2^{3/2}}$  whose minimum $s_2=s_1/2$ is weakly affected by presence of rare random loops.



\begin{figure*}[t]
    \centering
    \includegraphics[width=\linewidth]{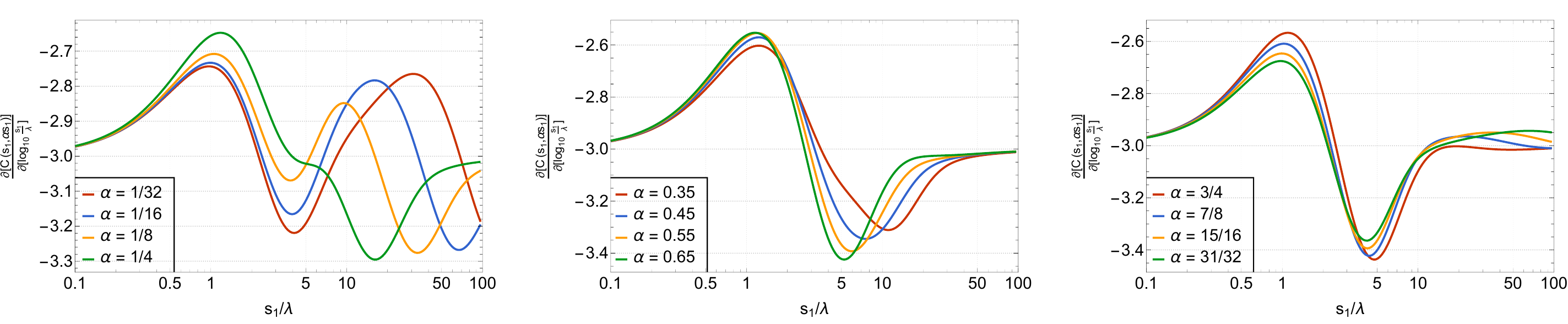}
    \caption{Log-log derivative $\frac{d \log _{10} C(s_1,\alpha s_1)}{d \log _{10} \frac{s_1}{\lambda}}$ (see Eq. (\ref{eq:log3})) in dependence on $s_1$ for different values of $\alpha$ at $\lambda/d=0.1$. 
    Note that for $\alpha\ll1$ the first argument of the pair correlation function is much larger than the second so that $C(s_1,\alpha s_1)\approx P(s_1)P(\alpha s_1)$. This explains the presence of the second maximum at $s_1\approx \lambda/\alpha$ in the left most panel.  
    }
    \label{Fig.4}
\end{figure*}

Equations (\ref{contact_prob_res}) and (\ref{eq:scaleless}) suggest that an appropriate graphical representation of the data on contact statistics may help to estimate the mean loop length.
Namely,  Fig. \ref{Fig: 5} demonstrates the  logarithmic derivative of the contact probability  
\begin{eqnarray}
    \frac{d \log _{10}P\left(s\right)}{d \log _{10} \frac{s}{\lambda}} \approx -\frac{3}{2}+\frac{\lambda}{d}s\frac{d}{ds}f(\frac{s}{\lambda}),
\end{eqnarray}
and the logarithmic derivative of the pair correlator for sufficiently small $s_2/\lambda$
 \begin{eqnarray}
      \label{eq:log1}
    &&\frac{d \log _{10}C\left(s_1, s_2\right)}{d \log _{10} \frac{s_1}{\lambda}} \approx -\frac{3s_1}{2(s_1-s_2)}+\\
    &&+\frac{\lambda}{d} \frac{s_1}{\lambda} F_{z_1}^{\prime}\left(\frac{s_1}{\lambda}, \frac{s_2}{\lambda}\right),
 \end{eqnarray}
attains maximum value at $s\sim \lambda$.
Analogous observation has been previously made in Ref. \cite{polovnikov2023crumpled} for log-log derivative of $P(s)$ based on non-perturbative semi-analytical calculations accounting all many-loop diagrams in our model. 
Note that for  $s_2\ll s_1$ the function $\frac{d \log _{10}C\left(s_1, s_2\right)}{d \log _{10} \frac{s_1}{\lambda}}$ approaches $\frac{d \log _{10}P\left(s_1\right)}{d \log _{10} \frac{s_1}{\lambda}}$ since in this case contacts become statistically independent from each other so that  $C(s_1,s_2)\approx P(s_1)P(s_2)$ (see also Fig. (\ref{Fig: 7}) below). 
Similar observations takes place for the logarithmic derivative of the pair correlation function 
 \begin{eqnarray}
      \label{eq:log2}
    &&\frac{d \log _{10}C\left(s_1, s_2\right)}{d \log _{10} \frac{s_2}{\lambda}} \approx -\frac{3}{2}+\frac{3s_2}{2(s_1-s_2)}+\\
    &&+\frac{\lambda}{d} \frac{s_2}{\lambda} F_{z_2}^{\prime}\left(\frac{s_1}{\lambda}, \frac{s_2}{\lambda}\right),
 \end{eqnarray}
in its dependence on $s_2$, see Fig. \ref{Fig: 6}.
When $s_1\gg s_2$, it approaches $\frac{d \log _{10}P\left(s_2\right)}{d \log _{10} \frac{s_2}{\lambda}}$.


\begin{figure}[t]
    \centering
    \includegraphics[scale=0.2]{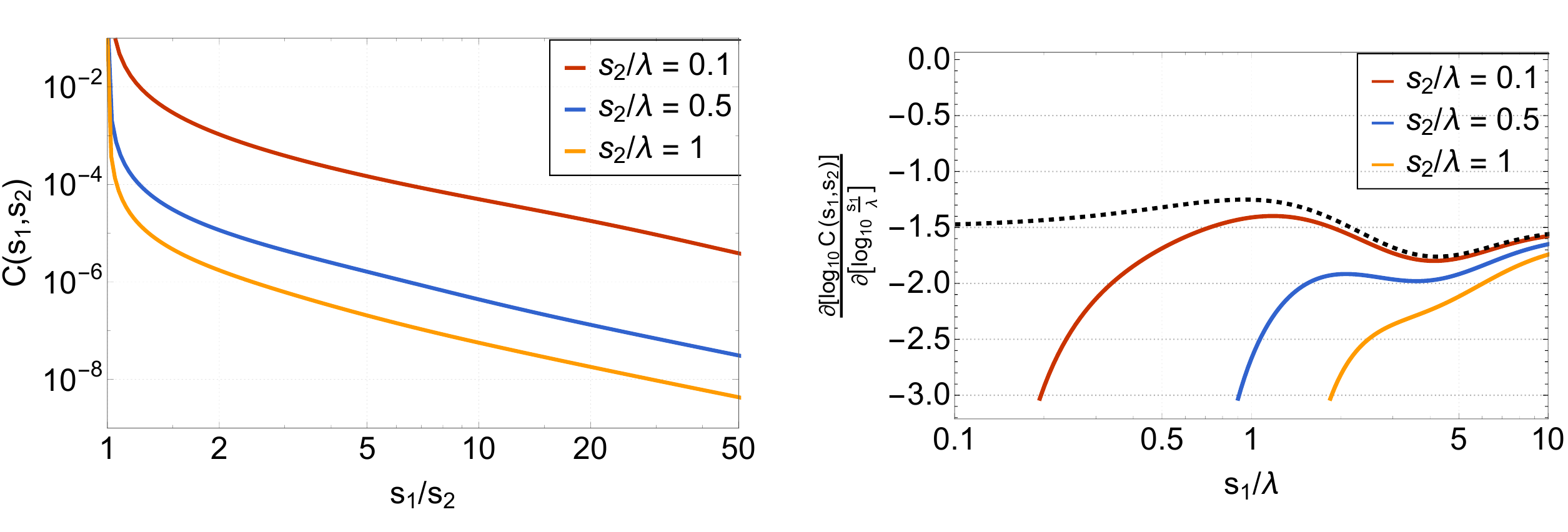}
    \caption{Left:  Pair correlation function $C(s_1,s_2)$ in dependence on $s_1/s_2$ at $\lambda/d=0.1$ for different values of $s_2/\lambda$.
    Right: Log-log derivative of pair-correlation function $C(s_1,s_2)$ (see Eq. (\ref{eq:scaleless})) in dependence on $s_1/\lambda$ at $\lambda/d=0.1$ for different values of $s_2/\lambda$.
    The dotted line corresponds to the log-log derivative of the contact probability $P(s_1)$ (see Eq. (\ref{contact_prob_res})).}
    \label{Fig: 5}
\end{figure}

Next let us pass to effectively one-dimensional  representation  of the pair correlator $C(s_1,s_2)$ by introducing the following function
    \begin{eqnarray}
        \label{eq:log3}
    \frac{d \log _{10} C(s_1,\alpha s_1)}{d \log _{10} \frac{s_1}{\lambda}}
    =-3 + \frac{\lambda}{d}s_1\frac{dF\left( \frac{s_1}{\lambda}, \frac{\alpha s_1}{\lambda}\right)}{ds_1},
    \end{eqnarray}
where $0<\alpha<1$ is some constant.
As shown in Fig. \ref{Fig.4} this object in its dependence on $s_1$ exhibits multiple extrema whose number depends on the value of parameter $\alpha$. 
Namely, there are four 
extrema at $\alpha<1/4$.
Next, in the case $1/4<\alpha<3/4$ we observe two extrema.
Finally, one has three extrema at $\alpha>3/4$.
Importantly, for any value of $\alpha$ there exists a maximum at $s\sim \lambda$.
Such rich behavior, if confirmed by analysis of the corresponding experimental data, would serve as another good confirmation of the effectiveness of our model.
Note, however, that for sufficiently small $\alpha$ the right-most extrema may not be observed experimentally since their positions correspond to large genomic scales (see first panel of Fig. \ref{Fig.4}) where contact statistics is affected by  compartmentalization \cite{nuebler2018chromatin}  - an effect which is not incorporated into the simple model considered here.


\begin{figure}[t]
    \centering
    \includegraphics[scale=0.2]{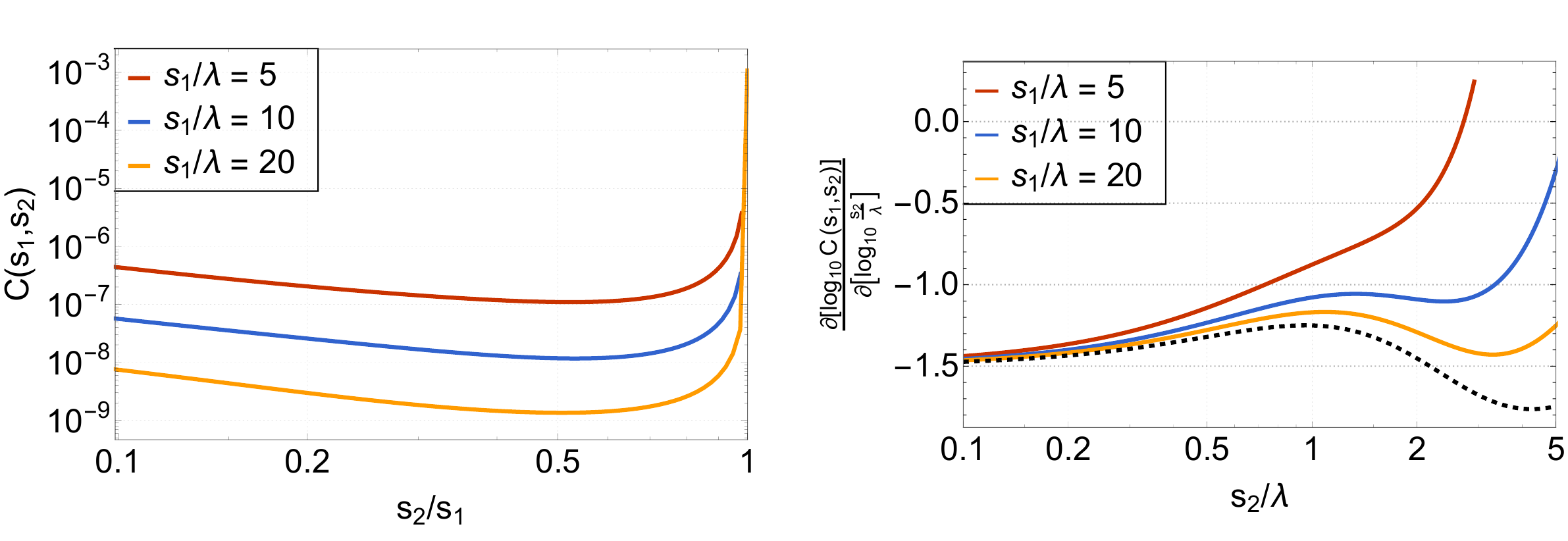}
    \caption{Left: Pair correlation function $C(s_1,s_2)$ in dependence on $s_2/s_1$ at $\lambda/d=0.1$ for different values of $s_1/\lambda$.  Right: Log-log derivative of pair-correlation function $C(s_1,s_2)$ (see Eq. (\ref{eq:scaleless})) in dependence on $s_2/\lambda$ at $\lambda/d=0.1$ for different values of $s_1/\lambda$.
    The dotted line corresponds to the log-log derivative of the contact probability $P(s_2)$ (see Eq. (\ref{contact_prob_res})).}
    \label{Fig: 6}
\end{figure}

\begin{figure}[t]
    \centering
    \includegraphics[scale=0.19]{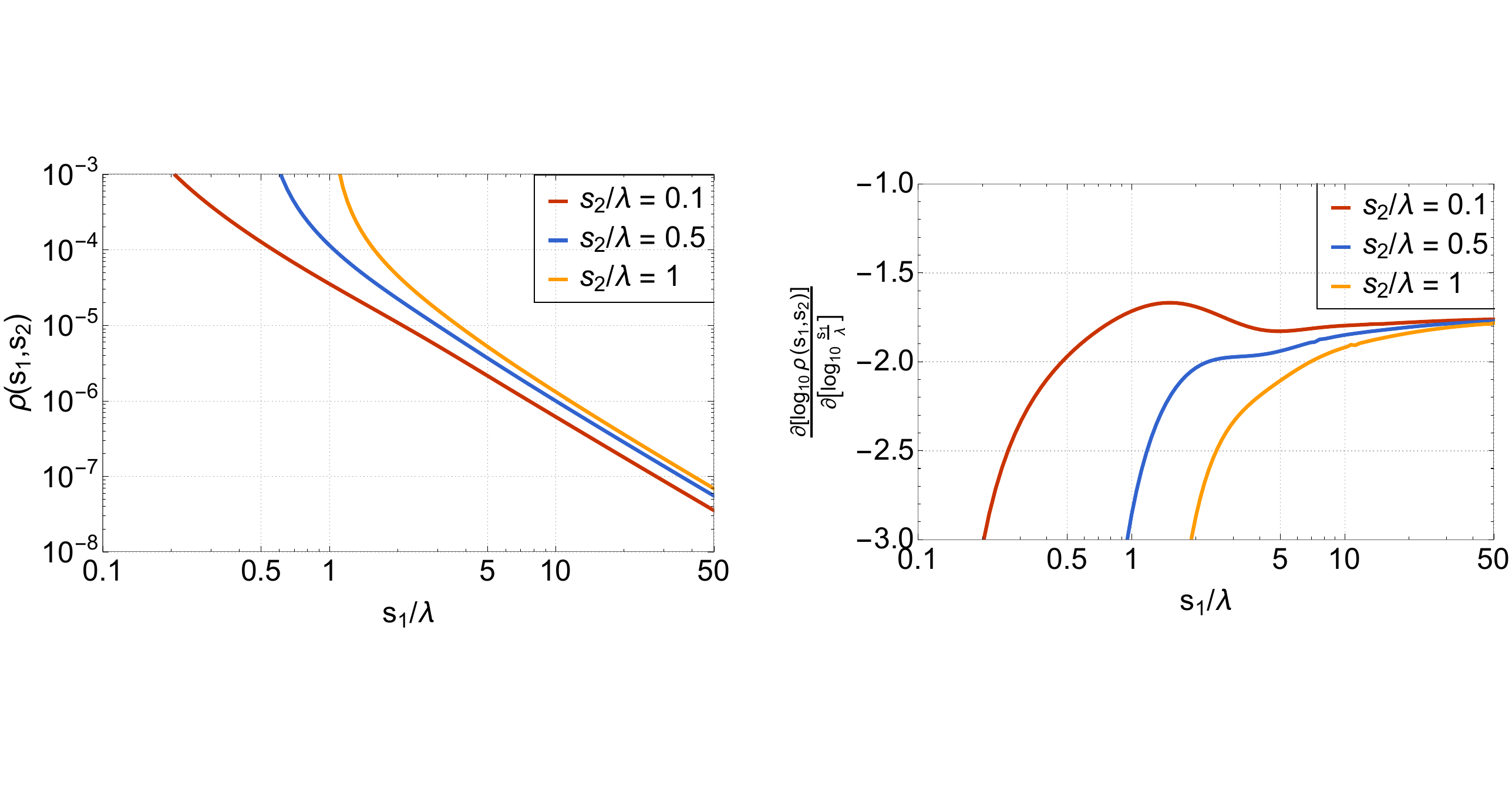}
    \caption{Left: The Pearson correlation coefficient $\rho(s_1,s_2)$ (see Eq. (\ref{Pearson_res})) of the random variables $\sigma_{i_1j_1}$ and $\sigma_{i_2j_2}$ in dependence on $s_1$ for different values of $s_2$ at $\lambda/d=0.1$. Right: The logarithmic derivative of the Pearson coefficient $\frac{d\log_{10}\rho(s_1,s_2)}{d\log_{10}\frac{s_1}{\lambda}}$ (see Eq. (\ref{log_Pearson_res})).}
    \label{Fig: 7}
\end{figure}



Finally, let us consider the Pearson correlation coefficient defined as
\begin{eqnarray}
    \rho=\frac{\langle (\sigma_{i_1j_1}-\langle \sigma_{i_1j_1}\rangle)(\sigma_{i_2j_2}-\langle \sigma_{i_2j_2}\rangle) \rangle}{\sqrt{\langle(\sigma_{i_1j_1}-\langle \sigma_{i_1j_1}\rangle)^2\rangle}\sqrt{\langle(\sigma_{i_2j_2}-\langle \sigma_{i_2j_2}\rangle)^2\rangle}},
\end{eqnarray}
to quantify linear correlation between random variables $\sigma_{i_1j_1}$ and $\sigma_{i_2j_2}$.
It is easy to check that $-1\le \rho\le +1$.

From Eqs. (\ref{contact_prob_res}) and (\ref{eq:scaleless}) we immediately find that 
\begin{eqnarray}
    \rho=\frac{C(s_1,s_2)-P(s_1)P(s_2)}{\sqrt{P(s_1)P(s_2)(1-P(s_1))(1-P(s_2))}}.
\end{eqnarray}
Next, taking into account that  
$P(s_1),P(s_2)\ll 1$ at genomic scales $s_1,s_2\gg a^2/l_{\text{eff}}$, in the leading order approximation in small parameters $s_1/d$, $s_2/d$, $b/s_1^{3/2}$, $b/s_2^{3/2}$ and $\lambda/d$ one obtains  
\begin{widetext}
\begin{eqnarray}
\label{Pearson_res}
    \rho (s_1,s_2)\approx 
   \frac{C(s_1,s_2)-P(s_1)P(s_2)}{\sqrt{P(s_1)P(s_2)}}\approx 
   b\frac{s_1^{3/2}-(s_1-s_2)^{3/2}}{s_1^{3/4}s_2^{3/4}(s_1-s_2)^{3/2}}\left(1+\frac{b}{2s_1^{3/2}}+\frac{b}{2s_2^{3/2}}+\frac{\lambda}{d}\Phi(\frac{s_1}{\lambda},\frac{s_2}{\lambda})\right),
\end{eqnarray}
and
\begin{eqnarray}
    \label{log_Pearson_res}\frac{d\log_{10}\rho(s_1,s_2)}{d\log_{10}\frac{s_1}{\lambda}}\approx -\frac{3}{4} + \frac{3}{2}\frac{s_1^{3/2}-s_1\cdot (s_1-s_2)^{1/2}}{s_1^{3/2}-(s_1-s_2)^{3/2}}-\frac{3s_1}{2(s_1-s_2)}-\frac{3b}{4s_1^{3/2}}+\frac{s_1}{\lambda} \frac{\lambda}{d}\Phi_{z_1}'\left(\frac{s_1}{\lambda},\frac{s_2}{\lambda} \right),
\end{eqnarray}
\end{widetext}
where
\begin{eqnarray}
&&\Phi(s_1,s_2)=-\frac{1}{2}(f(s_1)+f(s_2)) +\\+
    &&\frac{s_1^{3/2}F(s_1,s_2)-(s_1-s_2)^{3/2}(f(s_1)+f(s_2))}{s_1^{3/2}-(s_1-s_2)^{3/2}}.
\end{eqnarray}
Figure (\ref{Fig: 7}) illustrates Eqs. (\ref{Pearson_res}) and (\ref{log_Pearson_res}).
For sufficiently small $s_2$ the logarithmic derivative $\frac{d\log_{10}\rho(s_1,s_2)}{d\log_{10}\frac{s_1}{\lambda}}$ exhibits non-monotonic behaviour with maximum at $s_1\sim \lambda$ (see red curve in right panel of Fig. (\ref{Fig: 7})). 
Similar representation from $s_2$-perspective for fixed $s_1$ is not shown here as 
the corresponding dependence is  monotonic and thus it cannot be used to infer information regarding characteristic loop size.


\section{Conclusion}
The study of contact statistics of chromatin may shed light onto relationship between the spatial structure of the genome and its functioning \cite{hafner2022spatial}. The main regulatory mechanism that initiates gene expression in a living cell is the formation of chromatin loops, bringing enhancers and promoters into spatial proximity. Importantly, some enhancers and promoters are located at genomic distances of several hundred thousand base pairs from each other. To understand to what extent  loop extrusion may take a part in the regulation of enhancer-promoter communication in a cell of particular type one should be able to estimate the typical genomic size of cohesin-mediated loops.  At the moment, several predictors of the mean loop size have been proposed it the literature:  maximum of the log-log derivative of the contact probability \cite{polovnikov2023crumpled}, maximum of the log-log derivative of the conditional contact probability \cite{belan2022influence},  minimum of the log-log derivative of mean squared physical distance between genomic loci \cite{belan2023footprints},  maximum of kurtosis of the  separation vector \cite{belan2023footprints}. In this work we proposed several new predictors based on the behaviour of pair correlation function $C(s_1,s_2)$ and Pearson coefficient of the contact indicators associated with single-cell contact maps.


\acknowledgments

The work was supported by the Russian Science Foundation,
project no. 22-72-10052.



\appendix 

\begin{widetext}

\section{Derivation of Eq. (\ref{contact_prob_res})}
\label{app1}

\begin{center}
    \includegraphics[scale=0.3]{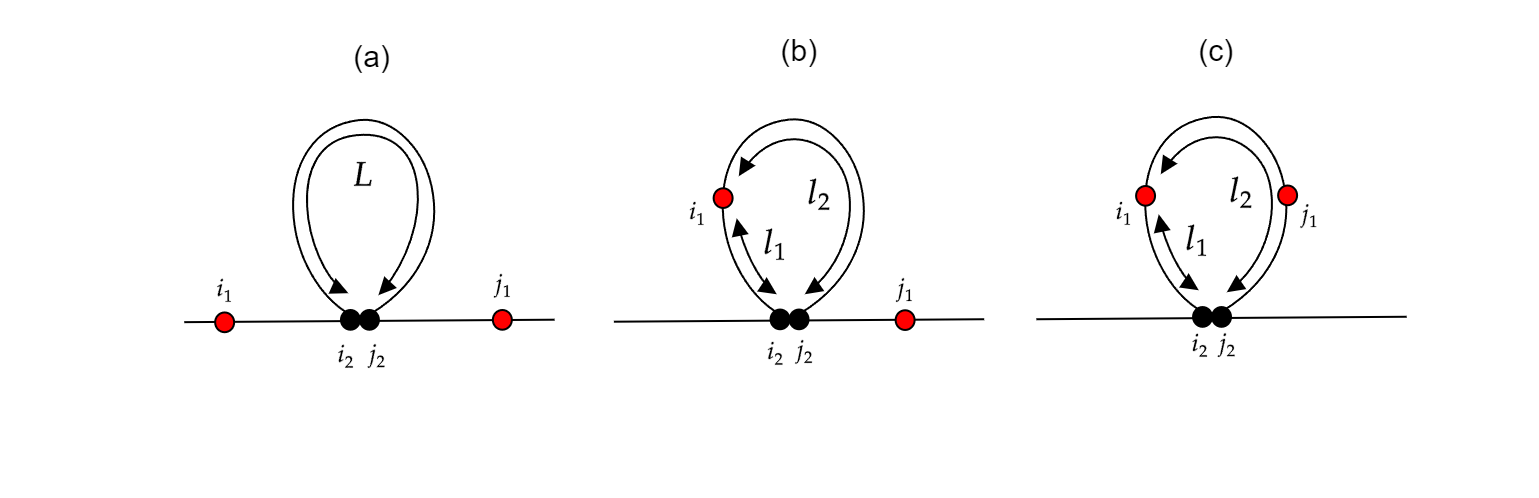}
    \label{Fig.app1}
\end{center}
Let us consider three one-loop diagrams contributing to the pairwise contact probability $P_c(s)$ in the rare loop limit, as shown in the figure above. Following results of \cite{polovnikov2023crumpled} we start with the expression
\begin{eqnarray}
\label{app_P_alpha}
&&{\cal P}_\alpha(0|\{A\}_\alpha;s)=\frac{1}{(\frac{2}{3}\pi l_{\text{eff}}\tilde s_\alpha[s,\{A\}_\alpha])^{3/2}},
\end{eqnarray}
where 
\begin{eqnarray}
\label{s_eff_a_one}
&\tilde{s}_{a}[s,x_s]=(1-x_s)s,&\\
\label{s_eff_b_one}
&\tilde{s}_{b}[s,l_1,l_2]=s-l_2+\frac{l_1l_2}{l_1+l_2},&\\
\label{s_eff_c_one}
&\tilde{s}_{c}[s,l_1,l_2]=\left(1-\frac{s}{l_1+l_2}\right)s.&
\end{eqnarray}
Here $x_s=L/s$ and the contour lengths $l_1$ and $l_2$ are shown in the figure above.

The linearized statistical wights of diagrams are given by
 \begin{eqnarray}
 \label{Wa}
&{\cal W}_{a}(x_s|s)\approx\delta(x_s)+\frac{\lambda}{d}\left(-(1+\frac{s}{\lambda})\delta(x_s)+\frac{(1-x_s)s^2}{\lambda}p_{\text{loop}}(x_ss)\right), \ \ \text{for} \ \ 0\le x_s< 1,& \\
 \label{Wb}
 &{\cal W}_{b}(L,q|s)\approx2\frac{\lambda}{d}\frac{L}{\lambda}p_{\text{loop}}(L), \ \ \text{for} \ \ 0\le q\le min[1,\frac{s}{L}], L\ge 0,&\\
  \label{Wc}
&{\cal W}_{c}(L,q|s)\approx\frac{\lambda}{d}\frac{L}{\lambda}p_{\text{loop}}(L),  \ \ \text{for} \ \  \frac{s}{L}\le q\le 1, L\ge s,&
 \end{eqnarray}
 where $q=l_2/L$.

 Substituting Eqs. (\ref{Q_alpha}) with ${\cal P}_\alpha(0|\{A\}_\alpha;s)$ given by Eqs. (\ref{app_P_alpha}) into Eq. (\ref{P_decomposition}) and using Eqs. (\ref{s_eff_a_one})-(\ref{Wc}) we find 
\begin{eqnarray}
&&P(s)=\frac{4}{3}\pi a^{3}\sum_{\alpha=a,b,c}\langle{\cal P}_\alpha(0|\{A\}_\alpha;s)\rangle_{\text{loops}}=\sqrt{\frac{6}{\pi}}\frac{a^3}{l^{3/2}_{\text{eff}}}\int\limits_0^1 \frac{{\cal W}_{a}(x_s|s)}{\tilde{s}_{a}[s,x_s]^{3/2}}dx_s + \\
&&+\sqrt{\frac{6}{\pi}}\frac{a^3}{l^{3/2}_{\text{eff}}}\int_0^\infty dL \int_0^{min[1,s/L]} dq  \frac{{\cal W}_b(L,q|s)}{\tilde{s}_{b}[s,l_1,l_2]^{3/2}} + \\
&&+\sqrt{\frac{6}{\pi}}\frac{a^3}{l^{3/2}_{\text{eff}}}\int_s^\infty dL \int_{s/L}^{1} dq \frac{{\cal W}_c(L,q|s)}{\tilde{s}_{c}[s,l_1,l_2]^{3/2}}=\sqrt{\frac{6}{\pi}}\frac{a^3}{l^{3/2}_{\text{eff}}s^{3/2}}\left[1+\frac{\lambda}{d}\left(-1-\frac{s}{\lambda}+\frac{s^2}{\lambda}\int\limits_0^{1}\frac{p_{\text{loop}}(ys)dy}{(1-y)^{1/2}}\right) \right]+\\
&&+\sqrt{\frac{6}{\pi}}\frac{a^3}{l^{3/2}_{\text{eff}}s^{3/2}}\frac{s^2}{d}\left[\int\limits_0^{1}\frac{p_{\text{loop}}(ys)y dy}{(1-y)^{1/2}} + \int\limits_0^{+\infty}\frac{(y+1)^{1/2}p_{\text{loop}}((y+1)s)dy}{y^{1/2}}\right]+\\
&&+\sqrt{\frac{6}{\pi}}\frac{a^3}{l^{3/2}_{\text{eff}}s^{3/2}}\frac{s^2}{d}\int\limits_0^{+\infty}\frac{(y+1)^{3/2}p_{\text{loop}}((y+1)s) dy}{y^{1/2}}=\\
&&=\frac{b}{s^{3/2}}\left(1+\frac{\lambda}{d}\left[\left(3\frac{s}{\lambda}+1\right)\left(2\sqrt{\frac{s}{\lambda}}D\left(\sqrt{\frac{s}{\lambda}}\right)-1\right)
+\sqrt{\pi}(\frac{s}{\lambda})^2e^{-\frac{s}{\lambda}}\left(2U(\frac{1}{2};2;\frac{s}{\lambda})+U(\frac{1}{2};3;\frac{s}{\lambda})\right)\right] \right).
\end{eqnarray}
This result coincides with Eq. (\ref{contact_prob_res}) in the main text.

\section{Derivation of Eq. (\ref{eq:scaleless})}
\label{sec:3}

We pass to calculation of the conditional probability densities ${\cal P}_{\beta}(\vec{R}_{i_1j_1} = 0, \vec{R}_{i_2j_2} = 0 | \vec{R}_{i_3j_3} = 0, \{B \}_{\beta}; s_1, s_2)$ and of the statistical weights $\omega_{\beta}( \{B\}_{\beta}, s_1,s_2)$ required to derive analytical expression for the pair correlator $C(s_1,s_2)$.
Since $i_2-i_1=j_1-j_2$ by assumption (see the main text), 
5 diagrams out of 14 depicted in Fig. \ref{Fig: 3} are mirror images of others, so it is enough to consider 9 scenarios. 
For the sake of brevity of notation, we will denote the spatial position of the points having contour coordinates $i_k$ and $j_k$ as $\vec i_k$ and $\vec j_k$, respectively. 
Also, let us denote the contact event $\vec{R}_{i_kj_k} = 0$ as $c_k$ and write shortly $p_{\beta}(c_1,c_2|c_3)$  instead of ${\cal P}_{\beta}(\vec{R}_{i_1j_1} = 0, \vec{R}_{i_2j_2} = 0 | \vec{R}_{i_3j_3} = 0, \{B \}_{\beta}; s_1, s_2)$.


\subsection{Diagram 1}
\label{diag:1}

\begin{center}
    \includegraphics[width=0.3\linewidth]{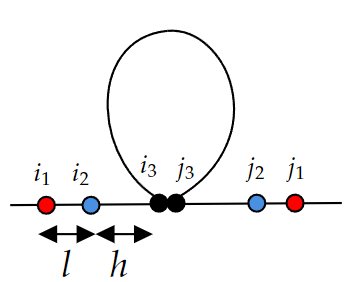}
    \label{Figdiag1}
\end{center}

First we consider diagram 1 in Fig. \ref{Fig: 3}.
Obviously
\begin{equation}
\label{factor: 2A}
    p_1(c_1,c_2|c_3) = \frac{\text{Pr}[c_1,c_2,c_3]}{\text{Pr}[c_3]} = \frac{\text{Pr}[c_1|c_2,c_3]\text{Pr}[c_2,c_3]} {\text{Pr}[c_3]} = \text{Pr}[c_1|c_2,c_3] \text{Pr}[c_2|c_3].
\end{equation}

The conditional probability of contact  $\text{Pr}[c_2|c_3]$ can be represented as 
\begin{equation}
    \label{c2c3}
    \text{Pr}[c_2|c_3] = \int_{R\leq a}  P(\vec{R}) d^3 \vec{R}\approx \frac{4}{3}\pi a^3 P(\vec{R} = 0),
\end{equation}
where $P(\vec{R})$ - is the probability density function of random vector $\vec{R}= (\vec{i_3} - \vec{i_2}) + (\vec{j_2} - \vec{j_3})$ which is equal to the sum of two independent normally distributed random vector. 
Recalling that the sum of independent Gaussian variables has normal statistics with the variance given by the sum of variances of the underlying terms and taking into account (\ref{pfree}) and (\ref{c2c3}), we obtain 
\begin{equation}
\label{length_reduction}
\text{Pr}[c_2|c_3] =  \frac{b}{(s_2 - L)^{3/2}}.
\end{equation}
Similarly it can be shown that 
\begin{equation}
\text{Pr}[c_1|c_2,c_3] =  \frac{b}{(s_1 - s_2)^{3/2}}.
\end{equation}
Thus, from Eq. (\ref{factor: 2A}) one obtains
\begin{equation}
\label{conditional:diag1}
    p_{1}(c_1,c_2|c_3) =  \frac{b^2}{(s_2 - L)^{3/2}(s_1 - s_2)^{3/2}}.
\end{equation}

For the statistical weight corresponding to the diagram 1 we find 
\begin{equation}
\begin{aligned}
    \omega_{1}(h, L|s_1,s_2, l) = \pi_{gap}p_{gap}(l+h)p_{loop}(L) \text{Pr}[t \geq s_1-l-h-L] =\\= \frac{d}{d + \lambda}\frac{1}{d}\exp(-\frac{l+h}{d}) \frac{1}{\lambda} \exp(-\frac{L}{\lambda})\int_{s_1-l-h-L}^{\infty}\frac{1}{d}\exp(-\frac{t}{d}) dt = \\
    \label{weight1}
    = \frac{1}{(d+\lambda)\lambda} \exp(-\frac{l+h}{d} -\frac{L}{\lambda}) \cdot \exp(-\frac{s_1-l-h-L}{d})
    ,
\end{aligned}
\end{equation}
where $l=i_2-i_1=j_1-j_2$, $h = i_3 - i_2$,  $\pi_{gap} = \frac{d}{d+\lambda}$ - is the probability that the 
point $i_1$ belongs to the gap region of the chain, 
$p_{gap}(l+h)$ - is the probability density to observe a gap of contour length $l+h$, $p_{loop}(L)$ - probability density of loop length $L$, $t$ -  is the random variable with probability density $p_{gap}(l)$, and $\text{Pr}[t \geq s_1-l-h-L]$ represents the probability that there are no additional loops between points $j_3$ and $j_1$.
When writing Eq. (\ref{weight1}), we took into account the  memoryless property of the exponentially distributed random variables.
Note that the random parameters $L$ and $h$ entering Eq. (\ref{weight1}) belong to the intervals $L \in [0,s_2]$ and $h \in [0,s_2-L]$, respectively.




\subsection{Diagram 2}
\label{diag:2}

\begin{center}
    \includegraphics[width=0.4\linewidth]{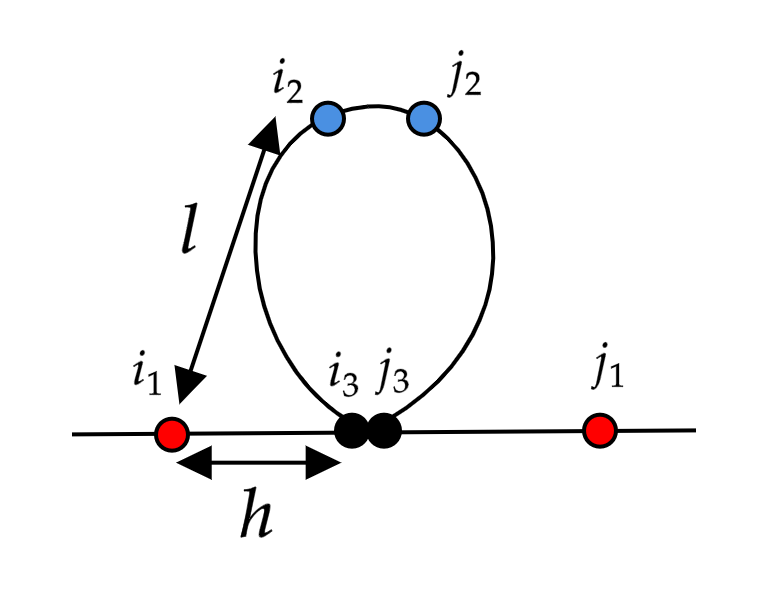}
    \label{Figdiag2}
\end{center}

The conditional probability associated with diagram 2 can be represented as
\begin{equation}
\label{diagram2_repr}
    p_2(c_1,c_2|c_3) = \text{Pr}[c_1|c_2,c_3] \text{Pr}[c_2|c_3].
\end{equation}
Following line of arguments similar to that presented in the previous subsection, we obtain 
\begin{equation}
\label{one_more_pr}
    \text{Pr}[c_1|c_2,c_3]=\frac{b}{(s_1 - L)^{3/2}}.
\end{equation}
Next, using Eq. (\ref{pcoil}) from the main text we obtain 
\begin{equation}
\label{conditional_bridge}
    \text{Pr}[c_2|c_3] = \frac{bL^{3/2}}{s_2^{3/2} \cdot (L - s_2)^{3/2}}.
\end{equation}
Substituting (\ref{one_more_pr}) and (\ref{conditional_bridge}) into (\ref{diagram2_repr}) we find 
\begin{equation}
\label{conditional:diag2}
    p_{2}(c_1,c_2|c_3) = \text{Pr}[c_1|c_2,c_3]  \text{Pr}[c_2|c_3] =  \frac{b^2L^{3/2}}{s_2^{3/2} \cdot (L - s_2)^{3/2} \cdot (s_1 - L)^{3/2}}.
\end{equation}

The statistical weight of the diagram 2 is given by
\begin{equation}
\begin{aligned}
    \omega_{2}(h, L|s_1,s_2, l) = \pi_{gap}p_{gap}(h)p_{loop}(L)\text{Pr}[t \geq s_1-h-L] = \\ =
    \frac{d}{d + \lambda}\frac{1}{d}\exp(-\frac{h}{d}) \frac{1}{\lambda} \exp(-\frac{L}{\lambda})\int_{s_1-h-L}^{\infty}\frac{1}{d}\exp(-\frac{t}{d}) =  \frac{1}{(d+\lambda)\lambda} \exp(-\frac{s_1-L}{d} -\frac{L}{\lambda}). 
\end{aligned}
\end{equation}
Now $L \in [s_2,s_1]$ and $h \in [Max[0,\frac{s_1+s_2}{2}-L],Min[\frac{s_1-s_2}{2},s_1-L] ]$. 
Such a choice guarantees that mutual positions of points $i_1,i_2,i_3,j_1,j_2$ and $j_3$ corresponds to situation depicted in Fig. \ref{Fig: 3}(2).




\subsection{Diagram 3}
\label{diag:3}

\begin{center}
    \includegraphics[width=0.4\linewidth]{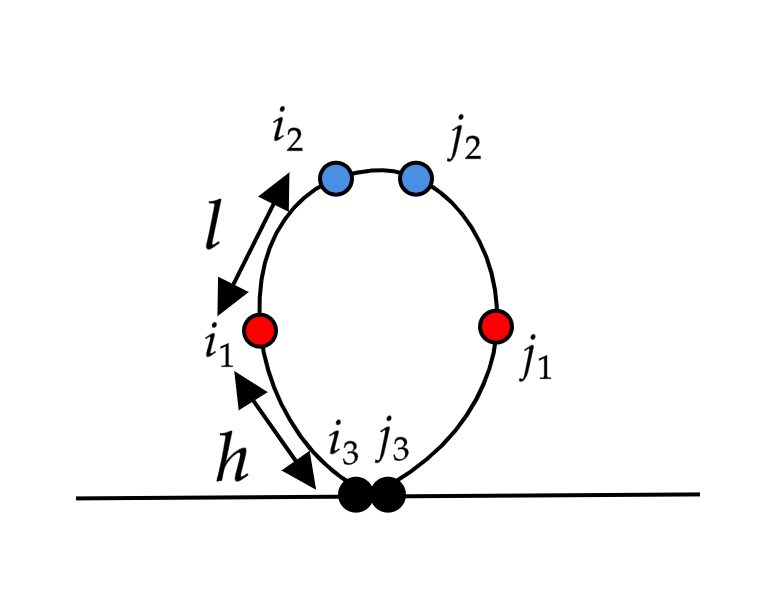}
    \label{Figdiag3}
\end{center}
As previously, let us use the following representation 
\begin{equation}
\label{conditional:diag3}
    p_3(c_1,c_2|c_3) = \text{Pr}[c_2|c_1,c_3] \text{Pr}[c_1|c_3].
\end{equation}
The second term in the right hand side of the above expression can be easily calculated using Eq. (\ref{pcoil}).
As for the first term, let us  note that  $\text{Pr}(c_2|c_1,c_3) = \text{Pr}(c_2|c_1)$ due to the Markovian property of the ideal chain. 
Given this fact, the corresponding term can also be calculated based on Eq. (\ref{pcoil}). 
This finally yields 
\begin{center}
\includegraphics[width=0.5\linewidth]{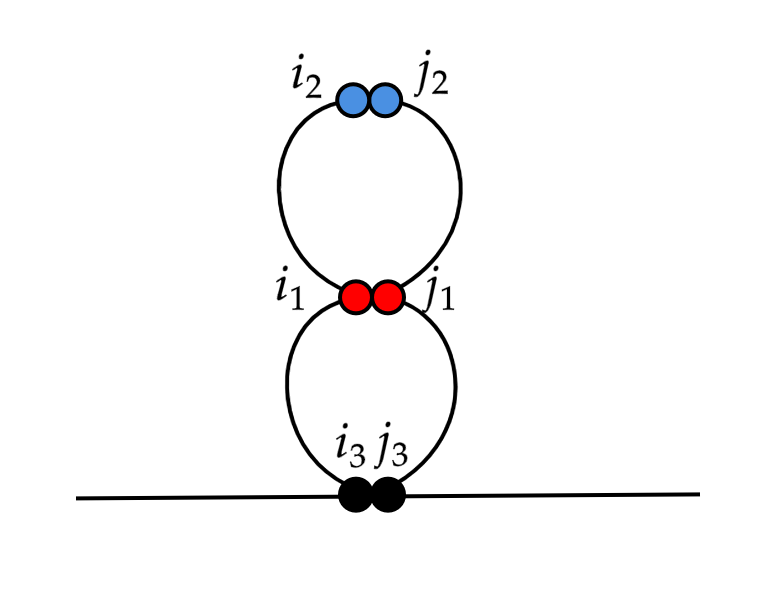}
    \label{Figdiag3_2}
\end{center}
\begin{equation}
\label{conditional:diag3_2}
     p_{3}(c_1,c_2|c_3) =  \frac{b^2L^{3/2}}{s_2^{3/2}  (s_1 - s_2)^{3/2} (L - s_1)^{3/2}}. 
\end{equation}


To arrive at the statistical weight of the diagram 3 let us introduce a variable $q = h/L$ defined as the ratio in which the point $i_1$ divides the cohesin-mediated loop.
In terms of variables $L$ and $q$ the statistical weight is given by 
\begin{equation}
     \omega_{3}(q, L|s_1,s_2, l) = \pi_{loop} \frac{L}{\lambda}p_{loop}(L) \Theta(q)\Theta(1-q) = \frac{L}{d + \lambda} \frac{1}{\lambda} e^{-\frac{L}{\lambda}} \Theta(q)\Theta(1-q),
\end{equation}
where $\pi_{loop} = \frac{\lambda}{d + \lambda}$  represents the probability that point $i_1$ lies at the loop, $\frac{L}{\lambda}p_{loop}(L)$ 
is the probability density of the
length of this loop, and $\rho(q) = \Theta(q)\Theta(1-q)$ is the probability density of $q$.
Clearly $L \in [0,\infty]$, while $h \in [0,L-s_1]$ and, thus, $q = h/L \in [0, 1 - s_1/L]$.

\subsection{Diagram 4}
\label{sec:diag4a}
\begin{center}
    \includegraphics[width=0.4\linewidth]{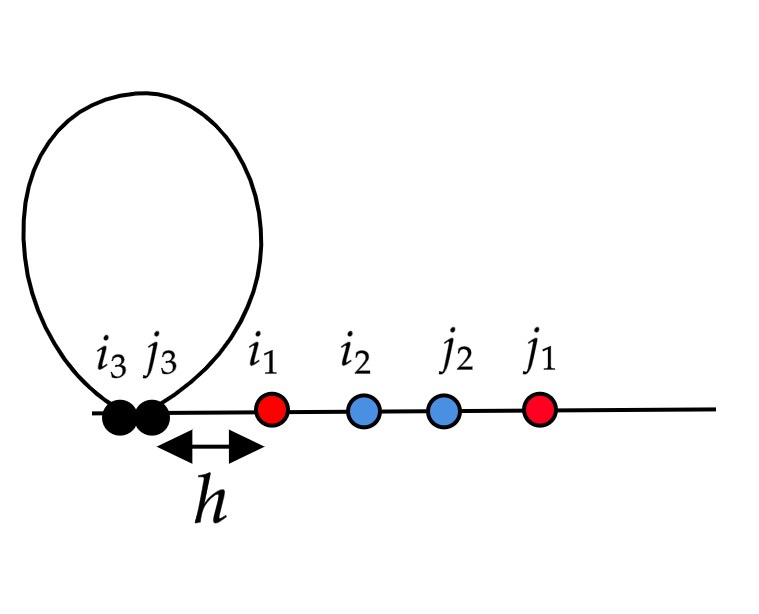}
    \label{Figdiag4}
\end{center}
For the conditional probability corresponding to the diagram 4 we find 
\begin{equation}
\label{conditional:diag4}
    p_{4}(c_1,c_2|c_3) =  \text{Pr}[c_1|c_2,c_3]  \text{Pr}[c_2|c_3] = \text{Pr}[c_1|c_2]  \text{Pr}[c_2]=\frac{b^2}{(s_1 - s_2)^{3/2}s_2^{3/2}},
\end{equation}
where we exploited the Markovian property of the ideal chain.
The probability of observing chain conformation describing by this diagram is given by 
\begin{equation}
    \omega_{4}(h, L|s_1,s_2) = \pi_{gap} \text{Pr}[t \geq s_1] = \frac{d}{d + \lambda} \exp (-\frac{s_1}{d}).
\end{equation}
\subsection{Diagrams 5A and 5B}
\label{diag:5}
\begin{center}
    \includegraphics[width=0.4\linewidth]{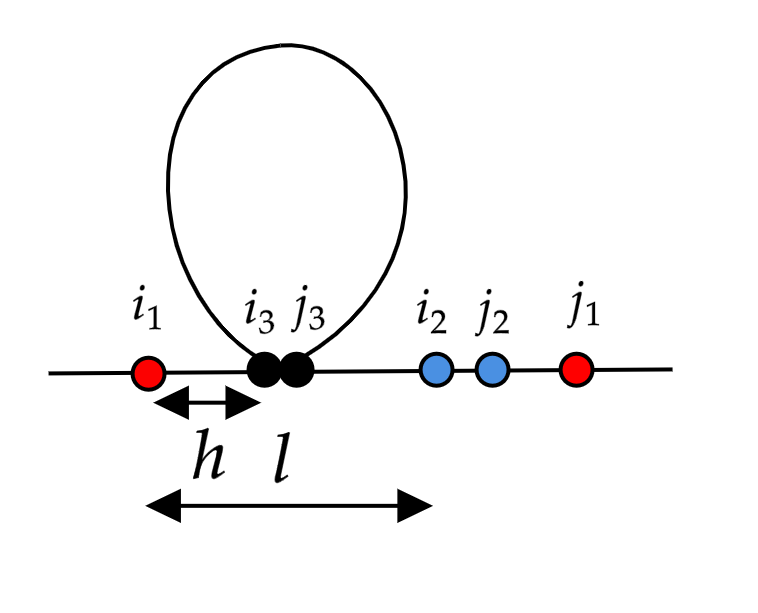}
    \label{Figdiag4}
\end{center}
Repeating arguments similar to those presented in (\ref{diag:1}), we find
\begin{equation}
\label{conditional:diag5}
    p_{5}(c_1,c_2|c_3) =   \text{Pr}[c_1|c_2,c_3]\text{Pr}[c_2|c_3] = \text{Pr}[c_1|c_2,c_3]\text{Pr}[c_2] =\frac{b^2}{(s_1 - s_2 - L)^{3/2}s_2^{3/2}}.
\end{equation}
The statistical weight of diagram 5 is as follows 
\begin{equation}
\begin{aligned}
    \omega_{5}(h, L|s_1,s_2, l) = \pi_{gap}p_{gap}(h)p_{loop}(L)\text{Pr}[t \geq s_1-h-L] = \\ =
    \frac{d}{d + \lambda}\frac{1}{d}\exp(-\frac{h}{d}) \frac{1}{\lambda} \exp(-\frac{L}{\lambda})\int_{s_1-h-L}^{\infty}\frac{1}{d}\exp(-\frac{t}{d})  = \frac{1}{(d+\lambda)\lambda} \exp(-\frac{s_1-L}{d} -\frac{L}{\lambda}). 
\end{aligned}
\end{equation}
The random variables $L$ and $h$ parametrising this diagram obey the conditions $L \in [0, l]$ and $h \in[0,l-L]$.


\begin{center}
    \includegraphics[width=0.4\linewidth]{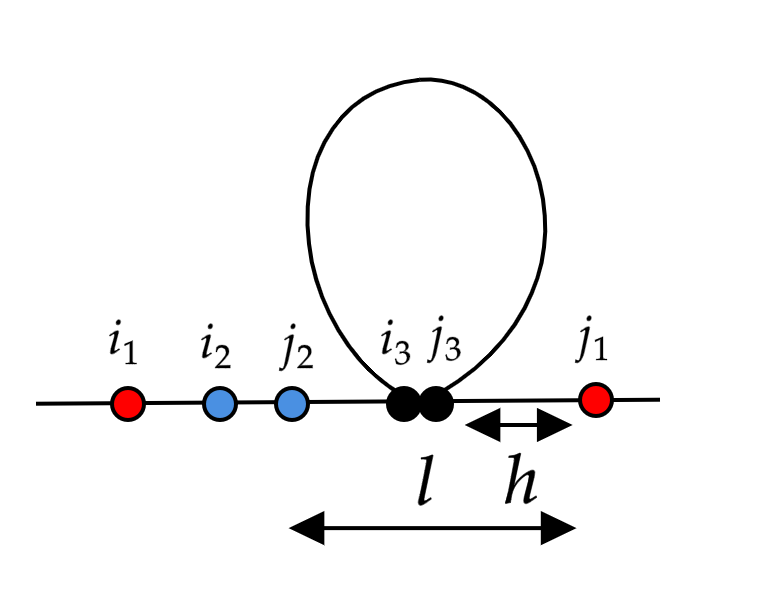}
    \label{Figdiag4}
\end{center}
Similarly for diagram 5B we find 
\begin{equation}
\label{diagram5B}
    p_{5B}(c_1,c_2|c_3) =  \text{Pr}[c_1|c_2,c_3] \text{Pr}[c_2|c_3] =  \frac{b^2}{(s_1 - s_2 - L)^{3/2}s_2^{3/2}},
\end{equation}
and
\begin{equation}
    \omega_{5B}(h, L|s_1,s_2, l) = \pi_{gap} p_{gap}[s_1-h-L] p_{loop}(L) \text{Pr}[t \geq h] = \frac{1}{(d+\lambda)\lambda} \exp(-\frac{s_1-L}{d} -\frac{L}{\lambda}),
\end{equation}
 where $L \in [0,l]$ and $h \in [0,l-L]$.

\subsection{Diagram 6A and 6B}
\label{diag:6}

\begin{center}
    \includegraphics[width=0.4\linewidth]{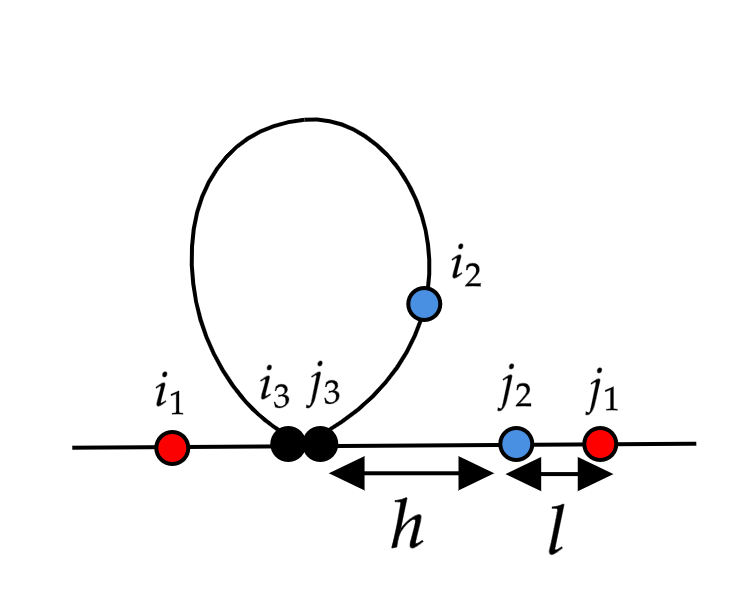}
    \label{Figdiag4}
\end{center}
We start again with representation 
\begin{equation}
\label{conditional:diag7}
    p(c_1,c_2|c_3) = \text{Pr}[c_2|c_1,c_3]  \text{Pr}[c_1|c_3].
\end{equation}
\begin{center}
    \includegraphics[width=0.5\linewidth]{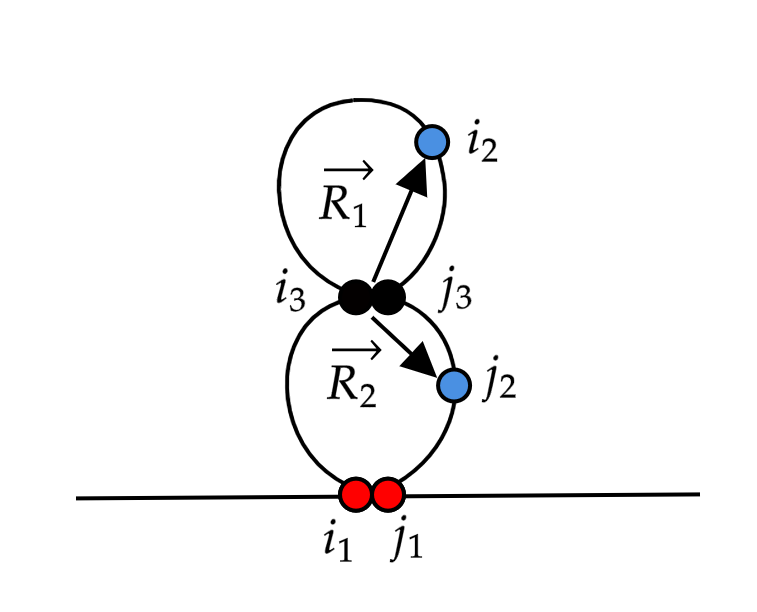}
    \label{Figdiag4}
\end{center}
The second term in the right hand side is similar to  (\ref{length_reduction}), 
so let us analyse the first one.
It is convenient to introduce the notations  $\vec R_1 = \vec i_2 - \vec j_3$ and $\vec R_2 = \vec j_2 - \vec j_3$.
The probability densities of the random vectors  $\vec{R_1}$ and $\vec{R_2}$ are given by normal distributions $P_{coil}(\vec R_1|j_3-i_2,s_3)$ and $P_{coil}(\vec R_2|j_2-j_3,s_1-s_3)$, respectively (see Eq. (\ref{pcoil})). 
Then the contact probability $\text{Pr}[c_2|c_1,c_3]$ can be found from the probability density of the vector $\vec{R} = \vec{R_2} - \vec{R_1}$.
Taking into account that variances of $\vec R_1$ and $\vec R_2$ behave as $\sigma_1^2 \sim \frac{(s_2-h)(L-s_2+h)}{L}$ and $\sigma_2^2 \sim \frac{h(s_1-L-h)}{s_1-L} $, we obtain 
\begin{equation}
\label{conditional:diag7}
\begin{aligned}
    p_{6}(c_1,c_2|c_3) = \frac{b^2}{(s_1 - L)^{3/2} \cdot \left( \frac{(s_2-h)(L-s_2+h)}{L} + \frac{h(s_1-L-h)}{s_1-L} \right)^{3/2}} = \\
    = \frac{b^2L^{3/2}}{\left[(s_1-L)(s_2-h)(L-s_2+h) + L\cdot h \cdot (s_1-L-h)\right]^{3/2}} = \frac{b^2L^{3/2}}{\left[s_2 \cdot (s_1 - L) \cdot (L+2h-s_2)-h^2s_1\right]^{3/2}}.
\end{aligned}
\end{equation}

For the statistical weight of the diagram 6 one obtains 
\begin{equation}
\begin{aligned}
    \omega_{6}(h, L|s_1,s_2, l) = \pi_{gap} p_{gap}(s_1-l-h-L) p_{loop}(L) \text{Pr}[t \geq (h+l)] = \\ = \frac{d}{\lambda + d} \frac{1}{d}\exp \left(-\frac{1}{d}(s_1-L) \right)\frac{1}{\lambda} \exp(-\frac{L}{\lambda}),
\end{aligned}
\end{equation}
where $L\in [0,s_1 - l]$ and $h \in [max(0,s_2-L),min(s_2,\frac{s_1+s_2}{2}-L)]$.

The non-trivial range for $h$ is explained by the need  to simultaneously satisfy three conditions: (i) the point $i_2$ should belong to the cohesin-mediated loop so that $h \leq s_2$ and $h \ge s_2 - L$, (ii) the total contour length of the internal segments should not exceed the contour distance between the external contacts, i.e. $h \leq s_1 - l - L$, (iii) the right point of internal contact lies to the right of the loop, i.e. $h\ge 0$.

\begin{center}
    \includegraphics[width=0.4\linewidth]{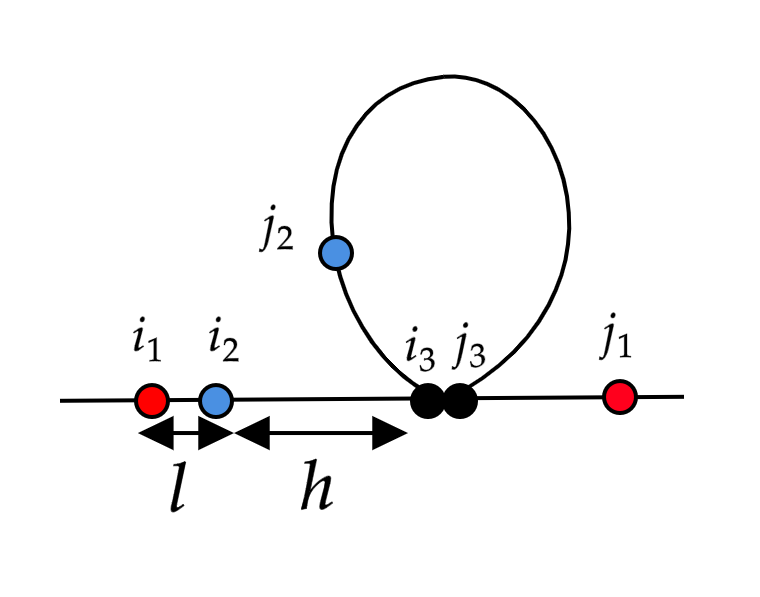}
    \label{Figdiag9}
\end{center}


\subsection{Diagrams 7A and 7B}
\label{diag:7}
\begin{center}
    \includegraphics[width=0.4\linewidth]{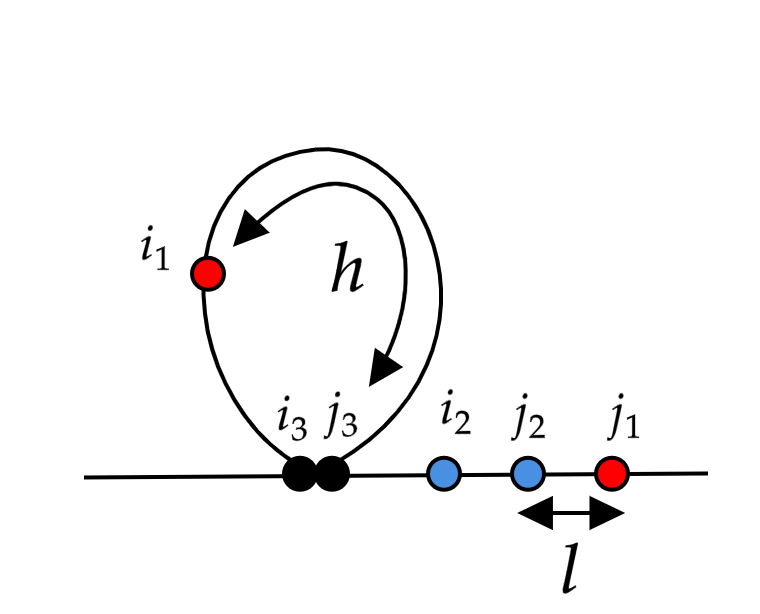}
    \label{Figdiag9}
\end{center}

\begin{equation}
    p_7(c_1,c_2|c_3) =  \text{Pr}[c_1|c_2,c_3] \text{Pr}[c_2|c_3]= \text{Pr}[c_1|c_2,c_3]\text{Pr}[c_2].
\end{equation}
From Eq. (\ref{pfree}) we immediately find 
\begin{eqnarray}
    \text{Pr}[c_2]=\frac{b^2}{s_2^{3/2}}.
\end{eqnarray}
To calculate $\text{Pr}[c_1|c_2,c_3]$ let us introduce notations  $\vec{R} = \vec{R_1} + \vec{R_2} = \vec{j_1} - \vec{i_1}$, $\vec{R_1} = \vec{j_1} - \vec{j_3}$ and $\vec{R_2} = \vec{j_3} - \vec{i_1}$.
\begin{center}
    \includegraphics[width=0.4\linewidth]{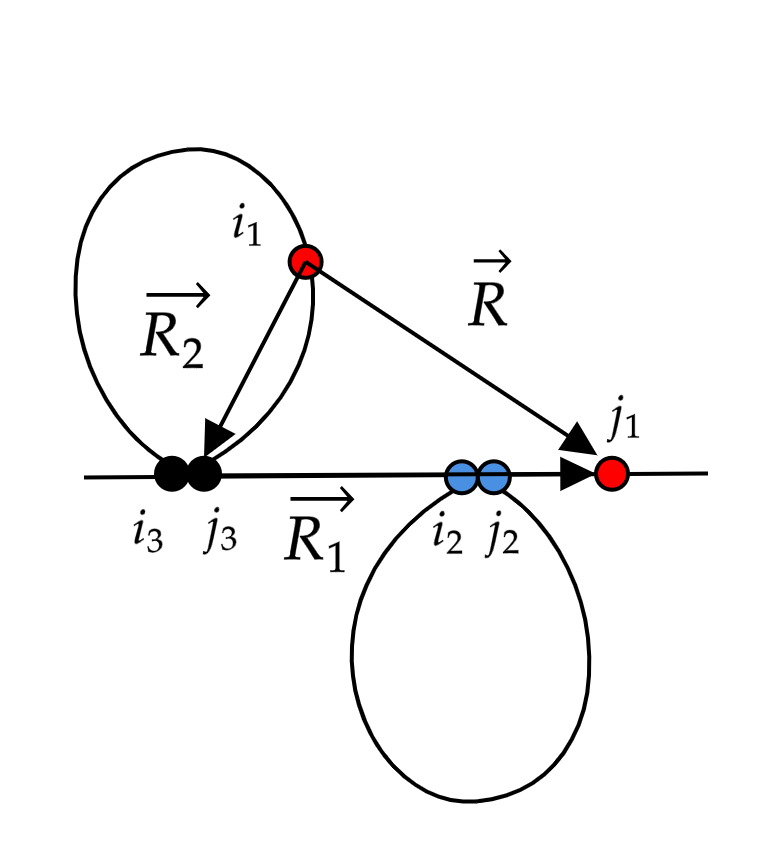}
    \label{Figdiag9}
\end{center}
The conditional contact probability $\text{Pr}[c_1|c_2,c_3]$  can be found from probability density of the random vector  $\vec{R} = \vec{R_1} + \vec{R_2}$.
Since independent vectors $\vec{R_1}$ and $\vec{R_2}$ have normal statistics with variances  $\sigma_1^2 \sim (j_1-j_3) - s_2 = s_1 - h - s_2  $ and $\sigma_2^2 \sim \frac{h(L-h)}{L}$, one obtains 
\begin{equation}
\label{conditional:diag7b}
    p_{7}(c_1,c_2|c_3) = \frac{b^2}{s_2^{3/2} \cdot \left(s_1-h-s_2+\frac{h(L-h)}{L} \right)^{3/2}} =  \frac{b^2L^{3/2}}{s_2^{3/2} \cdot \left(L(s_1-s_2) - h^2 \right)^{3/2}}.
\end{equation}


The statistical weight of the diagram 7 can be expressed as 
\begin{equation}
    \omega_{7}(q, L|s_1,s_2, l) = \pi_{loop} \frac{L}{\lambda}p_{loop}(L) \Theta(q)\Theta(1-q) \text{Pr}[t \geq s_1 - h],
\end{equation}
where $q = h/L$ and $h = j_3 - i_1$.
As follows from graphical representation, see Fig. \ref{Fig: 3}, the random variables $L$ and $h$ lie in the ranges $L \in [0,\infty)$ and $h \in [0,min(L,s_1-s_2-l)]$. 
Therefore, $h\in [0,min(1,\frac{s_1-s_2}{2L})]$.


\begin{center}
    \includegraphics[width=0.4\linewidth]{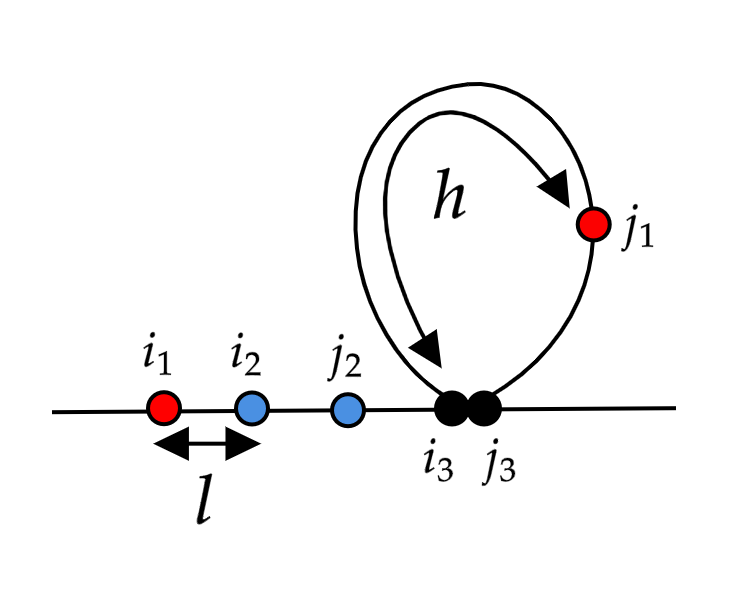}
    \label{Figdiag7breversed}
\end{center}


\subsection{Diagrams 8A and 8B}
\begin{center}
    \includegraphics[width=0.4\linewidth]{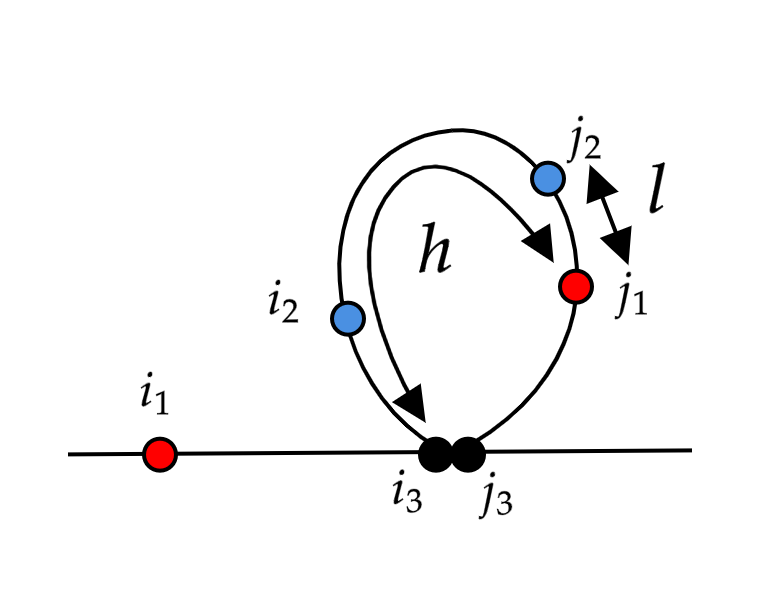}
    \label{Figdiag10a}
\end{center}
Let us use the representation 
\begin{equation}
\label{B28}
    p_8(c_1,c_2|c_3) = \text{Pr}[c_1|c_2,c_3] \text{Pr}[c_2|c_3].
\end{equation}
For the second term one readily obtains from Eq. (\ref{pcoil}) 
\begin{eqnarray}
\label{B29}
\text{Pr}[c_2|c_3] \sim \frac{L^{3/2}}{s_2^{3/2} \cdot(L-s_2)^{3/2}}.
\end{eqnarray}

To arrive at analytical expression for $\text{Pr}[c_1|c_2,c_3]$ let us introduce the vectors $\vec{R_1} = \vec{j_3} - \vec{i_1}$ and $\vec{R_2} = \vec{j_1} - \vec{j_3}$.
Then conditional probability $\text{Pr}[c_1|c_2,c_3]$ can be derived based on the probability density of random vector $\vec{R} = \vec{R_1} + \vec{R_2} = \vec{j_1} - \vec{i_1}$.
Since $\vec R_1$ and $\vec R_2$ have normal statistics with variances $\sigma_1^2 \sim (j_3-i_1) - s_3 = i_3 - i_1 = s_1-h$ and $\sigma_2^2 \sim \frac{(j_3-j_1)(s_3-s_2-j_3+j_1)}{s_3-s_2} = \frac{(s_3-h)(s_3-s_2-s_3+h)}{s_3-s_2} = \frac{(s_3-h)(h-s_2)}{s_3-s_2}$, we find
\begin{equation}
\label{B30}
\text{Pr}[c_1|c_2,c_3]\sim\frac{1}{\left(s_1 - h + \frac{(L-h)(h-s_2)}{L-s_2} \right)^{3/2}}.
\end{equation}
Inserting expressions (\ref{B29}) and (\ref{B30}) into Eq. (\ref{B28}), we obtain 
\begin{center}
    \includegraphics[width=0.4\linewidth]{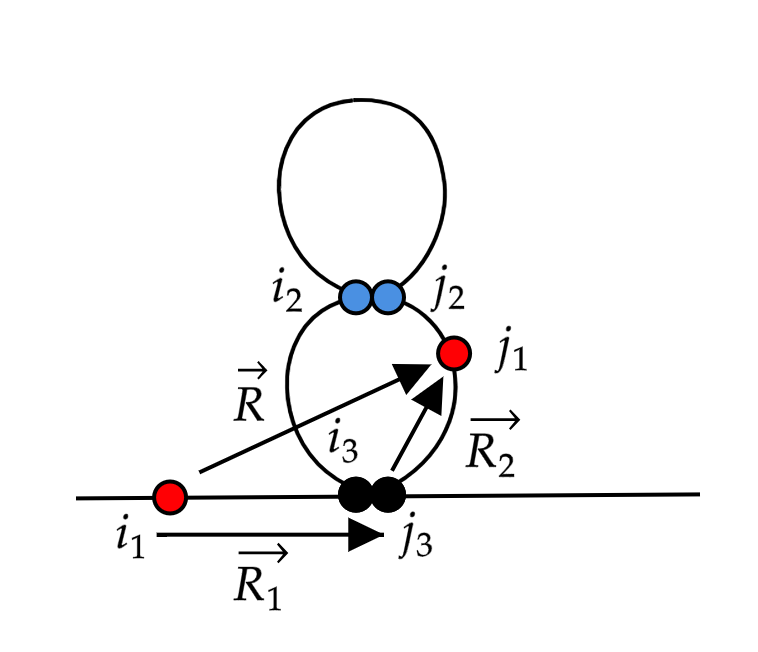}
    \label{Figdiag10a2}
\end{center}
\begin{equation}
\begin{aligned}
\label{conditional:diag10a}
    p^{(8)}(c_1,c_2|c_3) = A^2 \frac{L^{3/2}}{s_2^{3/2} \cdot(L-s_2)^{3/2}} \cdot \frac{1}{\left(s_1 - h + \frac{(L-h)(h-s_2)}{L-s_2} \right)^{3/2}} = \\ = A^2 \frac{L^{3/2}}{s_2^{3/2} \cdot \left((s_1-h)(L-s_2) + (L-h)(h-s_2))^{3/2}\right)}.
\end{aligned}
\end{equation}

The corresponding statistical weight equals to 
\begin{equation}
    \omega^{(8)}(h,L|s_1,s_2) = \pi_{gap} p_{gap}(s_1-h) p_{loop}(L),
\end{equation}
where $L \in [s_2+l, \infty)$and  $h \in[s_2 + l,min(L,s_1)]$.



\begin{center}
    \includegraphics[width=0.4\linewidth]{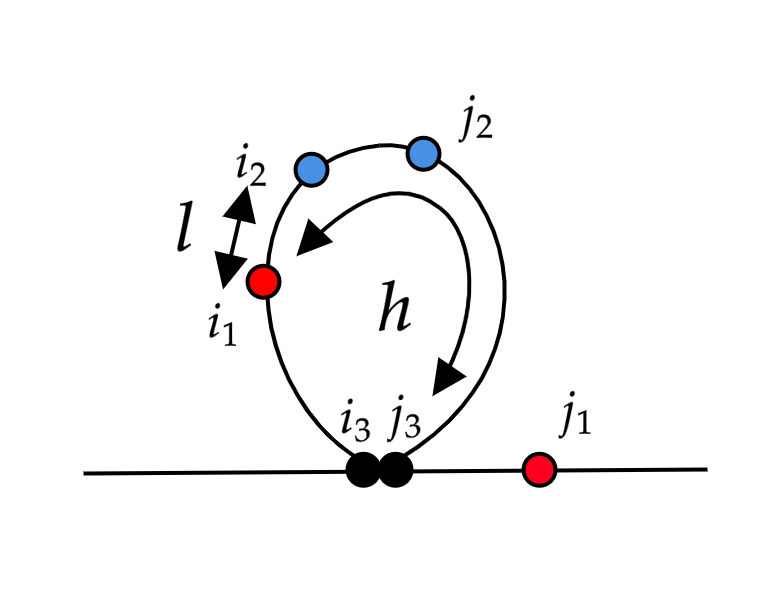}
    \label{Figdiag10c}
\end{center}


\subsection{Diagrams 9A and 9B}
\begin{center}
    \includegraphics[width=0.4\linewidth]{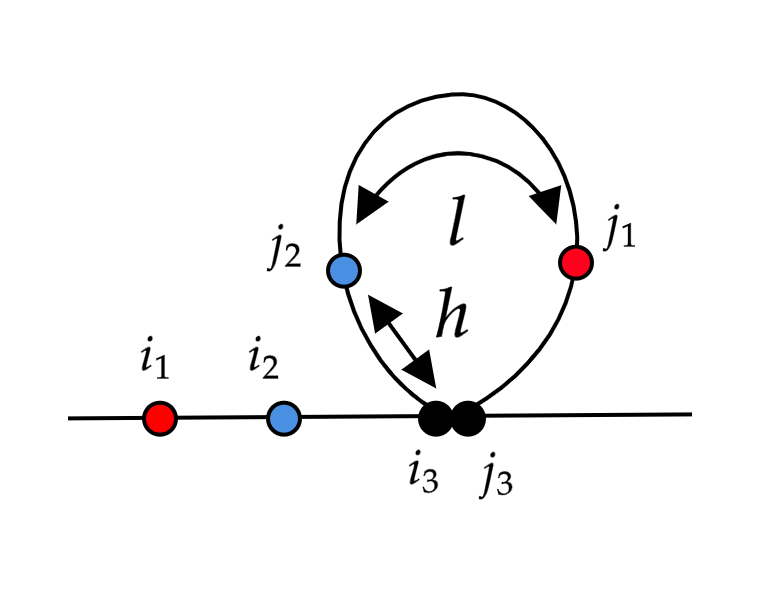}
    \label{Figdiag9b0}
\end{center}

\begin{center}
    \includegraphics[width=0.4\linewidth]{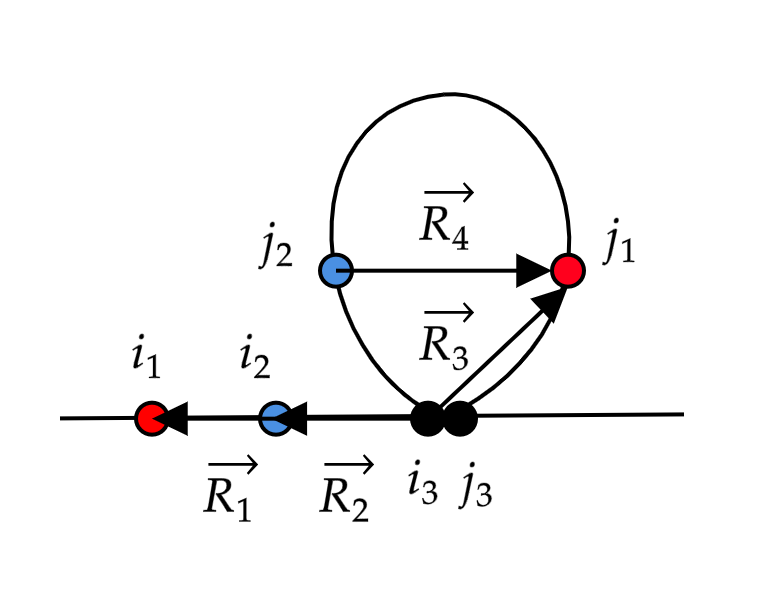}
    \label{Figdiag9b}
\end{center}

Le us introduce the notations  $\vec{r_1} = \vec{R_3} - \vec{R_2} - \vec{R_1}$ and $\vec{r_2} = -\vec{R_2} + \vec{R_3} - \vec{R_4}$.
The conditional probability  $p(c_1,c_2|c_3) = P(\vec{r_1} = 0, \vec{r_2} = 0 |c_3)$ can be calculated via averaging over realizations  of vectors $\vec{R_1}, \vec{R_2}, \vec{R_3}$ and $\vec{R_4}$.
Namely
\begin{equation}
\label{eq:long}
\begin{aligned}
    P(\vec{r_1},\vec{r_2} | c_3) = \langle \delta(\vec{r_1} - \vec{R_3} + \vec{R_1} + \vec{R_2}) \delta(\vec{r_2} + \vec{R_2} - \vec{R_3} + \vec{R_4})\rangle_{\vec{R_1},\vec{R_2},\vec{R_3},\vec{R_4}} = \\ =  \int d^3 R_1 d^3 R_2 d^3 R_3 d^3 R_4 \delta(\vec{r_1} - \vec{R_3} + \vec{R_1} + \vec{R_2}) \delta(\vec{r_2} + \vec{R_2} - \vec{R_3} + \vec{R_4}) P(\vec{R_1},\vec{R_2},\vec{R_3},\vec{R_4})= \\ =
    \int d^3 R_1 d^3 R_2 d^3 R_3 d^3 R_4 \delta(\vec{r_1} - \vec{R_3} + \vec{R_1} + \vec{R_2}) \delta(\vec{r_2} + \vec{R_2} - \vec{R_3} + \vec{R_4})P_{f}(\vec{R_1}|i_2-i_1) P_{f}(\vec{R_2}|i_3-i_2) P(\vec{R_3},\vec{R_4}),
\end{aligned}
\end{equation}
where we exploited the fact that  random vectors $\vec{R_1}$ and $\vec{R_2}$ are 
statistical independent from $\vec R_3$ and $\vec R_4$.
Note also that probability densities of  $\vec{R_1}$ and $\vec{R_2}$ are determined by Eq. \ref{pfree}). 

The joint probability density  $P(\vec{R_3}, \vec{R_4})$ cannot be factorized into the product of marginal probability densities  $P_{coil}(\vec{R_3}|s_3,L) \cdot P_{coil}(\vec{R_4}|s_3,L)$ since two vectors connecting pairs of point belonging the same loop are statistically correlated. 
Exploiting results of the work \cite{belan2022influence} we can write 
\begin{equation}
P\left(\vec{R_3}, \vec{R_4} \mid l, h, L\right)=\left(\frac{L}{4 \pi^2 h l\left(L - l - h\right)}\right)^{3 / 2} \exp \left(-\frac{\left(\vec{R_3}+\vec{R_4}\right)^2}{2 h}-\frac{\vec{R_3}^2}{2\left(L - l - h\right)}-\frac{\vec{R_4}^2}{2  l}\right),
\end{equation}
where $l = j_1 - j_2$ and $h = j_2-i_3$. 
There we can rewrite  Eq. (\ref{eq:long}) as
\begin{equation}
\begin{aligned}
p(c_1,c_2|c_3) = P(\vec{r_1} = 0, \vec{r_2} = 0|c_3) = \\ \int d^3 R_2 d^3 R_3 P_{f}(\vec{R_3} - \vec{R_2} | s_1-s_2-l) P_{f}(\vec{R_2}|s_2-h) P(\vec{R_3},\vec{R_3} - \vec{R_2}|l,h,L) \sim \\ \sim \frac{1}{(s_1-s_2-l)^{3/2}} \cdot \frac{1}{(s_2-h)^{3/2}} \cdot \left(\frac{L}{h l\left(L - l - h\right)}\right)^{3 / 2} \cdot \\ \cdot
\int d^3 R_2 d^3 R_3 \exp \left( - \frac{(\vec{R_3}-\vec{R_2})^2
}{2(s_1-s_2-l)} - \frac{\vec{R_2}^2}{2(s_2-h)} - \frac{\vec{R_2}^2}{2h} - \frac{(\vec{R_3}-\vec{R_2})^2
}{2l} - \frac{\vec{R_3}^2}{2(L-l-h)} -  \right) \sim \\ \sim \frac{1}{(s_1-s_2-l)^{3/2}} \cdot \frac{1}{(s_2-h)^{3/2}} \cdot \left(\frac{L}{h l\left(L - l - h\right)}\right)^{3 / 2} \cdot \frac{(L-l-h)^{3/2} \cdot (s_1-s_2-l)^{3/2} \cdot (lh)^{3/2} \cdot (s_2-h)^{3/2}}{\left((s_1-s_2)(Ls_2 - h^2) - l^2 s_2 \right)^{3/2}}. 
\end{aligned}
\end{equation}
Restoring the prefactor, we finally get
\begin{equation}
\label{conditional:diag9}
    p_{9}(c_1,c_2|c_3) = A^2 \frac{L^{3/2}}{\left((s_1-s_2)(Ls_2 - h^2) - l^2 s_2 \right)^{3/2}}.
\end{equation}

Finally, the probability of observing the diagram 9 is given by the following expression 
\begin{equation}
    \omega^{(9)}(h,L|s_1,s_2) = \pi_{gap}p_{gap}(s_1-l-h)p_{loop}(L),
\end{equation}
where $L \in [l,\infty) $ and $h \in [0, min (L-l,s_2)]$. 


Due to symmetry reasons, the diagram 9B, representing the mirror image of 9A, can be accounted by multiplying the contribution coming from 9A by factor $2$.
\begin{center}
    \includegraphics[width=0.4\linewidth]{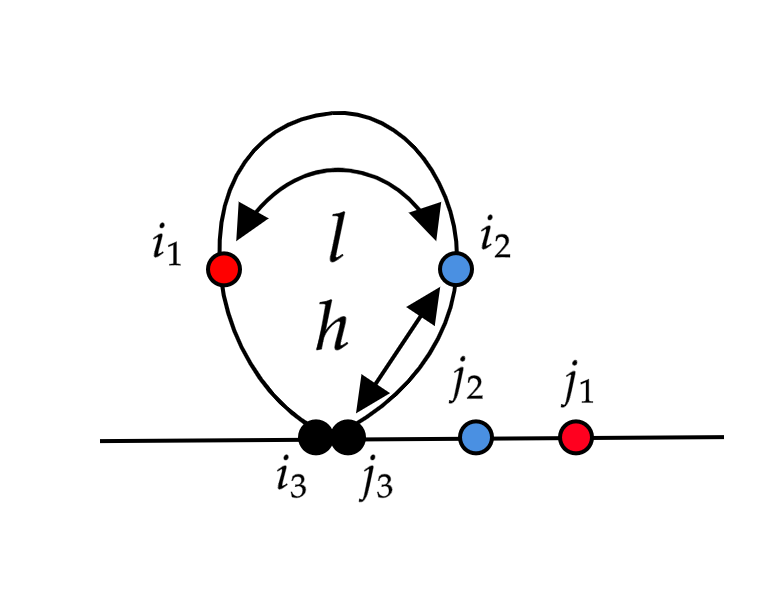}
    \label{Figdiag12a}
\end{center}

\section{Linearized statistical weights}
\label{sec: linearization}

As explained in the main text we restrict ourselves to one-loop diagrams.
Such an approximation is justified under assumption of  smallness of dimensionless parameters 
\begin{equation}
    \frac{\lambda}{d} \sim \frac{s_1}{d} \sim \frac{s_2}{d} \sim \varepsilon \ll 1.
\end{equation}

Since our calculations are correct only up to first order in small parameter $\varepsilon$, before calculating the pair correlator $C(s_1,s_2)$, the previously obtained expressions $\omega^{\alpha}(h,L|s_1,s_2)$ should be linearized
\begin{equation}
\begin{aligned}
\label{omega1}
    \omega^{(1)}(h,L|s_1,s_2) = \frac{1}{(d+\lambda)\lambda} \exp(-\frac{s_1-L}{d} -\frac{L}{\lambda}) = \frac{e^{-\frac{L}{\lambda}}}{\lambda d} \left(1 - \frac{s_1}{d} \right) \cdot \left(1 + \frac{L}{d}\right) \cdot \left(1 - \frac{\lambda}{d} \right) + O(\varepsilon^2) = \\ = \frac{e^{-\frac{L}{\lambda}}}{\lambda d} \left(1 - \frac{s_1 + \lambda - L}{d} \right) + O(\varepsilon^2),
\end{aligned}
\end{equation}
\begin{equation}
\begin{aligned}
\label{omega2}
    \omega^{(2)}(h,L|s_1,s_2) = \frac{1}{(d+\lambda)\lambda} \exp(-\frac{s_1-L}{d} -\frac{L}{\lambda}) = \frac{e^{-\frac{L}{\lambda}}}{\lambda d} \left(1 - \frac{s_1}{d} \right) \cdot \left(1 + \frac{L}{d}\right) \cdot \left(1 - \frac{\lambda}{d} \right) + O(\varepsilon^2) = \\ = \frac{e^{-\frac{L}{\lambda}}}{\lambda d} \left(1 - \frac{s_1 + \lambda - L}{d} \right) + O(\varepsilon^2),
\end{aligned}
\end{equation}
\begin{equation}
\begin{aligned}
\label{omega3}
    \omega^{(3)}(q, L|s_1,s_2, l) = \pi_{loop} \frac{L}{\lambda}p_{loop}(L) \Theta(q)\Theta(1-q) = \frac{L}{d + \lambda} \frac{1}{\lambda} e^{-\frac{L}{\lambda}} \Theta(q)\Theta(1-q) = \\ = \frac{Le^{-\frac{L}{\lambda}}}{\lambda d} \left(1 - \frac{\lambda}{d}\right)\Theta(q)\Theta(1-q) + O(\varepsilon^2),
\end{aligned}
\end{equation}
\begin{equation}
\label{omega4}
    \omega^{(4)}(h, L|s_1,s_2, l) =  \frac{d}{d + \lambda} \exp (-\frac{s_1}{d}) = \left(1 - \frac{s_1}{d} \right) \cdot \left(1 - \frac{\lambda}{d} \right) + O(\varepsilon^2) = 1 - \frac{s_1 + \lambda}{d} + O(\varepsilon^2)
\end{equation}
\begin{equation}
\begin{aligned}
\label{omega5}
    \omega^{(5)}(h,L|s_1,s_2) = \frac{1}{(d+\lambda)\lambda} \exp(-\frac{s_1-L}{d} -\frac{L}{\lambda}) = \frac{e^{-\frac{L}{\lambda}}}{\lambda d} \left(1 - \frac{s_1}{d} \right) \cdot \left(1 + \frac{L}{d}\right) \cdot \left(1 - \frac{\lambda}{d} \right) + O(\varepsilon^2) = \\ = \frac{e^{-\frac{L}{\lambda}}}{\lambda d} \left(1 - \frac{s_1 + \lambda - L}{d} \right) + O(\varepsilon^2),
\end{aligned}
\end{equation}
\begin{equation}
\begin{aligned}
\label{omega6}
    \omega^{(6)}(h,L|s_1,s_2) = \frac{1}{(d+\lambda)\lambda} \exp(-\frac{s_1-L}{d} -\frac{L}{\lambda}) = \frac{e^{-\frac{L}{\lambda}}}{\lambda d} \left(1 - \frac{s_1}{d} \right) \cdot \left(1 + \frac{L}{d}\right) \cdot \left(1 - \frac{\lambda}{d} \right) + O(\varepsilon^2) = \\ = \frac{e^{-\frac{L}{\lambda}}}{\lambda d} \left(1 - \frac{s_1 + \lambda - L}{d} \right) + O(\varepsilon^2),
\end{aligned}
\end{equation}
\begin{equation}
\begin{aligned}
\label{omega7}
    \omega^{(7)}(h,L|s_1,s_2) =\frac{L}{d + \lambda} \frac{1}{\lambda}\exp(-\frac{L}{\lambda} - \frac{s_1-qL}{d}) \Theta(q) \Theta(1-q) = \frac{Le^{- \frac{L}{\lambda}}}{\lambda d}  \left(1 - \frac{s_1 + \lambda -qL}{d} \right) \Theta(q)\Theta(1-q) + O(\varepsilon^2)
\end{aligned}
\end{equation}
\begin{equation}
\begin{aligned}
\label{omega8}
\omega^{(8)}(h,L|s_1,s_2) = \frac{1}{\lambda + d}\frac{1}{\lambda} \exp(-\frac{s_1-h}{d} - \frac{L}{\lambda}) = \frac{e^{-\frac{L}{\lambda}}}{\lambda d} \left(1 - \frac{s_1}{d} \right) \cdot \left(1 + \frac{h}{d}\right) \cdot \left(1 - \frac{\lambda}{d} \right) + O(\varepsilon^2) = \\ = \frac{e^{-\frac{L}{\lambda}}}{\lambda d} \left(1 - \frac{s_1 + \lambda - h}{d} \right) + O(\varepsilon^2).
\end{aligned}
\end{equation}
\begin{equation}
\begin{aligned}
\label{omega9}
\omega^{(9)}(h,L|s_1,s_2)\frac{1}{\lambda + d}\frac{1}{\lambda} \exp(-\frac{\frac{s_1+s_2}{2}-h}{d} - \frac{L}{\lambda}) = \frac{e^{-\frac{L}{\lambda}}}{\lambda d} \left(1 - \frac{s_1+s_2}{2d} \right) \cdot \left(1 + \frac{h}{d}\right) \cdot \left(1 - \frac{\lambda}{d} \right) + O(\varepsilon^2) = \\ = \frac{e^{-\frac{L}{\lambda}}}{\lambda d} \left(1 - \frac{s_1 + s_2 + 2\lambda - 2h}{2d} \right) + O(\varepsilon^2).
\end{aligned}
\end{equation}

\newpage

\section{Pair correlation function}


Multiplying conditional probabilities of contacts by linearized statistical weights and integrating over random variables associated with loops disorder according to Eq. (\ref{C_representation}) we arrive at Eq. (\ref{eq:scaleless})
where
\begin{eqnarray}
    \label{F_appendix}
    F(z_1,z_2) = -1 - z_1 + z_2^2 \int_0^1 d\tilde{L} \frac{e^{-\tilde{L}z_2} }{(1-\tilde{L})^{1/2}} + \\ + z_2^2\int_{1}^{\frac{1}{2}(1+z_1/z_2)}d\tilde{L} \frac{\tilde{L}^{3/2}  }{(\tilde{L}-1)^{1/2}} \cdot \frac{1}{\left(1 - \frac{z_2(\tilde{L}-1)}{z_1-z_2}\right)^{3/2}} \cdot e^{-\tilde{L}z_2}  + \\ +
    z_1^2\int_{\frac{1}{2}(1+z_2/z_1)}^{1}d\tilde{L} \frac{\tilde{L}^{3/2}  }{(1-\tilde{L})^{1/2}} \cdot \frac{1}{\left(1 - \frac{z_1(1-\tilde{L})}{z_1-z_2} \right)^{3/2}} \cdot e^{-\tilde{L}z_1}  + \\ +
    z_1^2\int_{1}^{\infty} d \tilde{L} \frac{\tilde{L}^{3/2}}{(\tilde{L} - 1)^{1/2}} \cdot e^{-\tilde{L}z_1} + \\ + \frac{(z_1-z_2)^2}{2}\int_{0}^{1} d \tilde{L} \frac{(1-\tilde{L})}{(1 -\tilde{L}/2)^{3/2}} \cdot e^{-\frac{\tilde{L}}{2}(z_1-z_2)} + \\ +
    2z_2^2 \int_0^1d\tilde{h} \int_1^{\frac{z_1+z_2}{2z_2}}d\tilde{L} \frac{(L-h)^{3/2} \cdot \left( 1- \frac{z_2}{z_1} \right)^{3/2} \cdot e^{-z_2(L-h)}}{\left[ (L+h-1)(1 - \frac{z_2}{z_1}(L-h)) - h^2 \right]^{3/2}}   + \\ +
    2(z_1-z_2)^2 \cdot\int_0^{1/2}d\tilde{L} \int_0^{1}dq \frac{\tilde{L}}{(1-q^2\tilde{L})^{3/2}} \cdot e^{-\tilde{L}(z_1-z_2)} + \\ +
    2(z_1-z_2)^2 \cdot\int_{1/2}^{\infty}d\tilde{L} \int_0^{\frac{1}{2\tilde{L}}}dq \frac{\tilde{L}}{(1-q^2\tilde{L})^{3/2}} \cdot e^{-\tilde{L}(z_1-z_2)}  + \\ +
    \frac{(z_1+z_2)^{2}}{2}\int_{1}^{\infty}d\tilde{L}\int_{1}^{min(\tilde{L},\frac{2z_1}{z_1+z_2})}d\tilde{h} \frac{\tilde{L}^{3/2} \cdot e^{-\frac{\tilde{L}}{2}(z_1+z_2)} }{\left(\tilde{L} + \tilde{h}\frac{2z_2}{z_1-z_2}-\tilde{h}^2\frac{z_1+z_2}{2(z_1-z_2)}-\frac{2z_1z_2}{z_1^2-z_2^2}\right)^{3/2}} + \\
    \label{F12}
    +
    \frac{(z_1-z_2)^2}{2}\int_1^{\infty}d\tilde{L}\int_0^{min(\tilde{L}-1,\frac{2z_2}{z_1-z_2})}d\tilde{h}\frac{\tilde{L}^{3/2 }\cdot e^{-\tilde{L}\frac{z_1-z_2}{2}}}{(\tilde{L}-\tilde{h}^2\cdot\frac{z_1-z_2}{2z_2}-\frac{1}{2})^{3/2}} + O(\varepsilon),
\end{eqnarray}
and we passed to dimensionless variables of integration.


\end{widetext}

\normalsize

\bibliography{main.bib}

\begin{thebibliography}{39}%
\makeatletter
\providecommand \@ifxundefined [1]{%
 \@ifx{#1\undefined}
}%
\providecommand \@ifnum [1]{%
 \ifnum #1\expandafter \@firstoftwo
 \else \expandafter \@secondoftwo
 \fi
}%
\providecommand \@ifx [1]{%
 \ifx #1\expandafter \@firstoftwo
 \else \expandafter \@secondoftwo
 \fi
}%
\providecommand \natexlab [1]{#1}%
\providecommand \enquote  [1]{``#1''}%
\providecommand \bibnamefont  [1]{#1}%
\providecommand \bibfnamefont [1]{#1}%
\providecommand \citenamefont [1]{#1}%
\providecommand \href@noop [0]{\@secondoftwo}%
\providecommand \href [0]{\begingroup \@sanitize@url \@href}%
\providecommand \@href[1]{\@@startlink{#1}\@@href}%
\providecommand \@@href[1]{\endgroup#1\@@endlink}%
\providecommand \@sanitize@url [0]{\catcode `\\12\catcode `\$12\catcode
  `\&12\catcode `\#12\catcode `\^12\catcode `\_12\catcode `\%12\relax}%
\providecommand \@@startlink[1]{}%
\providecommand \@@endlink[0]{}%
\providecommand \url  [0]{\begingroup\@sanitize@url \@url }%
\providecommand \@url [1]{\endgroup\@href {#1}{\urlprefix }}%
\providecommand \urlprefix  [0]{URL }%
\providecommand \Eprint [0]{\href }%
\providecommand \doibase [0]{https://doi.org/}%
\providecommand \selectlanguage [0]{\@gobble}%
\providecommand \bibinfo  [0]{\@secondoftwo}%
\providecommand \bibfield  [0]{\@secondoftwo}%
\providecommand \translation [1]{[#1]}%
\providecommand \BibitemOpen [0]{}%
\providecommand \bibitemStop [0]{}%
\providecommand \bibitemNoStop [0]{.\EOS\space}%
\providecommand \EOS [0]{\spacefactor3000\relax}%
\providecommand \BibitemShut  [1]{\csname bibitem#1\endcsname}%
\let\auto@bib@innerbib\@empty
\bibitem [{\citenamefont {Ganji}\ \emph {et~al.}(2018)\citenamefont {Ganji},
  \citenamefont {Shaltiel}, \citenamefont {Bisht}, \citenamefont {Kim},
  \citenamefont {Kalichava}, \citenamefont {Haering},\ and\ \citenamefont
  {Dekker}}]{ganji2018real}%
  \BibitemOpen
  \bibfield  {author} {\bibinfo {author} {\bibfnamefont {M.}~\bibnamefont
  {Ganji}}, \bibinfo {author} {\bibfnamefont {I.~A.}\ \bibnamefont {Shaltiel}},
  \bibinfo {author} {\bibfnamefont {S.}~\bibnamefont {Bisht}}, \bibinfo
  {author} {\bibfnamefont {E.}~\bibnamefont {Kim}}, \bibinfo {author}
  {\bibfnamefont {A.}~\bibnamefont {Kalichava}}, \bibinfo {author}
  {\bibfnamefont {C.~H.}\ \bibnamefont {Haering}},\ and\ \bibinfo {author}
  {\bibfnamefont {C.}~\bibnamefont {Dekker}},\ }\bibfield  {title} {\bibinfo
  {title} {Real-time imaging of dna loop extrusion by condensin},\ }\href@noop
  {} {\bibfield  {journal} {\bibinfo  {journal} {Science}\ }\textbf {\bibinfo
  {volume} {360}},\ \bibinfo {pages} {102} (\bibinfo {year}
  {2018})}\BibitemShut {NoStop}%
\bibitem [{\citenamefont {Golfier}\ \emph {et~al.}(2020)\citenamefont
  {Golfier}, \citenamefont {Quail}, \citenamefont {Kimura},\ and\ \citenamefont
  {Brugu{\'e}s}}]{golfier2020cohesin}%
  \BibitemOpen
  \bibfield  {author} {\bibinfo {author} {\bibfnamefont {S.}~\bibnamefont
  {Golfier}}, \bibinfo {author} {\bibfnamefont {T.}~\bibnamefont {Quail}},
  \bibinfo {author} {\bibfnamefont {H.}~\bibnamefont {Kimura}},\ and\ \bibinfo
  {author} {\bibfnamefont {J.}~\bibnamefont {Brugu{\'e}s}},\ }\bibfield
  {title} {\bibinfo {title} {Cohesin and condensin extrude dna loops in a cell
  cycle-dependent manner},\ }\href@noop {} {\bibfield  {journal} {\bibinfo
  {journal} {Elife}\ }\textbf {\bibinfo {volume} {9}},\ \bibinfo {pages}
  {e53885} (\bibinfo {year} {2020})}\BibitemShut {NoStop}%
\bibitem [{\citenamefont {Kong}\ \emph {et~al.}(2020)\citenamefont {Kong},
  \citenamefont {Cutts}, \citenamefont {Pan}, \citenamefont {Beuron},
  \citenamefont {Kaliyappan}, \citenamefont {Xue}, \citenamefont {Morris},
  \citenamefont {Musacchio}, \citenamefont {Vannini},\ and\ \citenamefont
  {Greene}}]{kong2020human}%
  \BibitemOpen
  \bibfield  {author} {\bibinfo {author} {\bibfnamefont {M.}~\bibnamefont
  {Kong}}, \bibinfo {author} {\bibfnamefont {E.~E.}\ \bibnamefont {Cutts}},
  \bibinfo {author} {\bibfnamefont {D.}~\bibnamefont {Pan}}, \bibinfo {author}
  {\bibfnamefont {F.}~\bibnamefont {Beuron}}, \bibinfo {author} {\bibfnamefont
  {T.}~\bibnamefont {Kaliyappan}}, \bibinfo {author} {\bibfnamefont
  {C.}~\bibnamefont {Xue}}, \bibinfo {author} {\bibfnamefont {E.~P.}\
  \bibnamefont {Morris}}, \bibinfo {author} {\bibfnamefont {A.}~\bibnamefont
  {Musacchio}}, \bibinfo {author} {\bibfnamefont {A.}~\bibnamefont {Vannini}},\
  and\ \bibinfo {author} {\bibfnamefont {E.~C.}\ \bibnamefont {Greene}},\
  }\bibfield  {title} {\bibinfo {title} {Human condensin i and ii drive
  extensive atp-dependent compaction of nucleosome-bound dna},\ }\href@noop {}
  {\bibfield  {journal} {\bibinfo  {journal} {Molecular cell}\ }\textbf
  {\bibinfo {volume} {79}},\ \bibinfo {pages} {99} (\bibinfo {year}
  {2020})}\BibitemShut {NoStop}%
\bibitem [{\citenamefont {Davidson}\ \emph {et~al.}(2019)\citenamefont
  {Davidson}, \citenamefont {Bauer}, \citenamefont {Goetz}, \citenamefont
  {Tang}, \citenamefont {Wutz},\ and\ \citenamefont
  {Peters}}]{davidson2019dna}%
  \BibitemOpen
  \bibfield  {author} {\bibinfo {author} {\bibfnamefont {I.~F.}\ \bibnamefont
  {Davidson}}, \bibinfo {author} {\bibfnamefont {B.}~\bibnamefont {Bauer}},
  \bibinfo {author} {\bibfnamefont {D.}~\bibnamefont {Goetz}}, \bibinfo
  {author} {\bibfnamefont {W.}~\bibnamefont {Tang}}, \bibinfo {author}
  {\bibfnamefont {G.}~\bibnamefont {Wutz}},\ and\ \bibinfo {author}
  {\bibfnamefont {J.-M.}\ \bibnamefont {Peters}},\ }\bibfield  {title}
  {\bibinfo {title} {Dna loop extrusion by human cohesin},\ }\href@noop {}
  {\bibfield  {journal} {\bibinfo  {journal} {Science}\ }\textbf {\bibinfo
  {volume} {366}},\ \bibinfo {pages} {1338} (\bibinfo {year}
  {2019})}\BibitemShut {NoStop}%
\bibitem [{\citenamefont {Kim}\ \emph {et~al.}(2019)\citenamefont {Kim},
  \citenamefont {Shi}, \citenamefont {Zhang}, \citenamefont {Finkelstein},\
  and\ \citenamefont {Yu}}]{kim2019human}%
  \BibitemOpen
  \bibfield  {author} {\bibinfo {author} {\bibfnamefont {Y.}~\bibnamefont
  {Kim}}, \bibinfo {author} {\bibfnamefont {Z.}~\bibnamefont {Shi}}, \bibinfo
  {author} {\bibfnamefont {H.}~\bibnamefont {Zhang}}, \bibinfo {author}
  {\bibfnamefont {I.~J.}\ \bibnamefont {Finkelstein}},\ and\ \bibinfo {author}
  {\bibfnamefont {H.}~\bibnamefont {Yu}},\ }\bibfield  {title} {\bibinfo
  {title} {Human cohesin compacts dna by loop extrusion},\ }\href@noop {}
  {\bibfield  {journal} {\bibinfo  {journal} {Science}\ }\textbf {\bibinfo
  {volume} {366}},\ \bibinfo {pages} {1345} (\bibinfo {year}
  {2019})}\BibitemShut {NoStop}%
\bibitem [{\citenamefont {Ryu}\ \emph {et~al.}(2020)\citenamefont {Ryu},
  \citenamefont {Katan}, \citenamefont {van~der Sluis}, \citenamefont {Wisse},
  \citenamefont {de~Groot}, \citenamefont {Haering},\ and\ \citenamefont
  {Dekker}}]{ryu2020condensin}%
  \BibitemOpen
  \bibfield  {author} {\bibinfo {author} {\bibfnamefont {J.-K.}\ \bibnamefont
  {Ryu}}, \bibinfo {author} {\bibfnamefont {A.~J.}\ \bibnamefont {Katan}},
  \bibinfo {author} {\bibfnamefont {E.~O.}\ \bibnamefont {van~der Sluis}},
  \bibinfo {author} {\bibfnamefont {T.}~\bibnamefont {Wisse}}, \bibinfo
  {author} {\bibfnamefont {R.}~\bibnamefont {de~Groot}}, \bibinfo {author}
  {\bibfnamefont {C.~H.}\ \bibnamefont {Haering}},\ and\ \bibinfo {author}
  {\bibfnamefont {C.}~\bibnamefont {Dekker}},\ }\bibfield  {title} {\bibinfo
  {title} {The condensin holocomplex cycles dynamically between open and
  collapsed states},\ }\href@noop {} {\bibfield  {journal} {\bibinfo  {journal}
  {Nature Structural \& Molecular Biology}\ }\textbf {\bibinfo {volume} {27}},\
  \bibinfo {pages} {1134} (\bibinfo {year} {2020})}\BibitemShut {NoStop}%
\bibitem [{\citenamefont {Banigan}\ and\ \citenamefont
  {Mirny}(2020)}]{banigan2020loop}%
  \BibitemOpen
  \bibfield  {author} {\bibinfo {author} {\bibfnamefont {E.~J.}\ \bibnamefont
  {Banigan}}\ and\ \bibinfo {author} {\bibfnamefont {L.~A.}\ \bibnamefont
  {Mirny}},\ }\bibfield  {title} {\bibinfo {title} {Loop extrusion: theory
  meets single-molecule experiments},\ }\href@noop {} {\bibfield  {journal}
  {\bibinfo  {journal} {Current opinion in cell biology}\ }\textbf {\bibinfo
  {volume} {64}},\ \bibinfo {pages} {124} (\bibinfo {year} {2020})}\BibitemShut
  {NoStop}%
\bibitem [{\citenamefont {Kimura}\ \emph {et~al.}(1999)\citenamefont {Kimura},
  \citenamefont {Rybenkov}, \citenamefont {Crisona}, \citenamefont {Hirano},\
  and\ \citenamefont {Cozzarelli}}]{kimura199913s}%
  \BibitemOpen
  \bibfield  {author} {\bibinfo {author} {\bibfnamefont {K.}~\bibnamefont
  {Kimura}}, \bibinfo {author} {\bibfnamefont {V.~V.}\ \bibnamefont
  {Rybenkov}}, \bibinfo {author} {\bibfnamefont {N.~J.}\ \bibnamefont
  {Crisona}}, \bibinfo {author} {\bibfnamefont {T.}~\bibnamefont {Hirano}},\
  and\ \bibinfo {author} {\bibfnamefont {N.~R.}\ \bibnamefont {Cozzarelli}},\
  }\bibfield  {title} {\bibinfo {title} {13s condensin actively reconfigures
  dna by introducing global positive writhe: implications for chromosome
  condensation},\ }\href@noop {} {\bibfield  {journal} {\bibinfo  {journal}
  {Cell}\ }\textbf {\bibinfo {volume} {98}},\ \bibinfo {pages} {239} (\bibinfo
  {year} {1999})}\BibitemShut {NoStop}%
\bibitem [{\citenamefont {Nasmyth}(2001)}]{nasmyth2001disseminating}%
  \BibitemOpen
  \bibfield  {author} {\bibinfo {author} {\bibfnamefont {K.}~\bibnamefont
  {Nasmyth}},\ }\bibfield  {title} {\bibinfo {title} {Disseminating the genome:
  joining, resolving, and separating sister chromatids during mitosis and
  meiosis},\ }\href@noop {} {\bibfield  {journal} {\bibinfo  {journal} {Annual
  review of genetics}\ }\textbf {\bibinfo {volume} {35}},\ \bibinfo {pages}
  {673} (\bibinfo {year} {2001})}\BibitemShut {NoStop}%
\bibitem [{\citenamefont {Riggs}(1990)}]{riggs1990dna}%
  \BibitemOpen
  \bibfield  {author} {\bibinfo {author} {\bibfnamefont {A.}~\bibnamefont
  {Riggs}},\ }\bibfield  {title} {\bibinfo {title} {Dna methylation and late
  replication probably aid cell memory, and type i dna reeling could aid
  chromosome folding and enhancer function},\ }\href@noop {} {\bibfield
  {journal} {\bibinfo  {journal} {Philosophical Transactions of the Royal
  Society of London. B, Biological Sciences}\ }\textbf {\bibinfo {volume}
  {326}},\ \bibinfo {pages} {285} (\bibinfo {year} {1990})}\BibitemShut
  {NoStop}%
\bibitem [{\citenamefont {Polovnikov}\ \emph {et~al.}(2023)\citenamefont
  {Polovnikov}, \citenamefont {Brand{\~a}o}, \citenamefont {Belan},
  \citenamefont {Slavov}, \citenamefont {Imakaev},\ and\ \citenamefont
  {Mirny}}]{polovnikov2023crumpled}%
  \BibitemOpen
  \bibfield  {author} {\bibinfo {author} {\bibfnamefont {K.~E.}\ \bibnamefont
  {Polovnikov}}, \bibinfo {author} {\bibfnamefont {H.~B.}\ \bibnamefont
  {Brand{\~a}o}}, \bibinfo {author} {\bibfnamefont {S.}~\bibnamefont {Belan}},
  \bibinfo {author} {\bibfnamefont {B.}~\bibnamefont {Slavov}}, \bibinfo
  {author} {\bibfnamefont {M.}~\bibnamefont {Imakaev}},\ and\ \bibinfo {author}
  {\bibfnamefont {L.~A.}\ \bibnamefont {Mirny}},\ }\bibfield  {title} {\bibinfo
  {title} {Crumpled polymer with loops recapitulates key features of chromosome
  organization},\ }\href@noop {} {\bibfield  {journal} {\bibinfo  {journal}
  {Physical Review X}\ }\textbf {\bibinfo {volume} {13}},\ \bibinfo {pages}
  {041029} (\bibinfo {year} {2023})}\BibitemShut {NoStop}%
\bibitem [{\citenamefont {Belan}\ and\ \citenamefont
  {Starkov}(2022)}]{belan2022influence}%
  \BibitemOpen
  \bibfield  {author} {\bibinfo {author} {\bibfnamefont {S.}~\bibnamefont
  {Belan}}\ and\ \bibinfo {author} {\bibfnamefont {D.}~\bibnamefont
  {Starkov}},\ }\bibfield  {title} {\bibinfo {title} {Influence of active loop
  extrusion on the statistics of triple contacts in the model of interphase
  chromosomes},\ }\href@noop {} {\bibfield  {journal} {\bibinfo  {journal}
  {JETP Letters}\ }\textbf {\bibinfo {volume} {115}},\ \bibinfo {pages} {763}
  (\bibinfo {year} {2022})}\BibitemShut {NoStop}%
\bibitem [{\citenamefont {Belan}\ and\ \citenamefont
  {Parfenyev}(2023)}]{belan2023footprints}%
  \BibitemOpen
  \bibfield  {author} {\bibinfo {author} {\bibfnamefont {S.}~\bibnamefont
  {Belan}}\ and\ \bibinfo {author} {\bibfnamefont {V.}~\bibnamefont
  {Parfenyev}},\ }\bibfield  {title} {\bibinfo {title} {Footprints of loop
  extrusion in statistics of intra-chromosomal distances: an analytically
  solvable model},\ }\href@noop {} {\bibfield  {journal} {\bibinfo  {journal}
  {arXiv preprint arXiv:2301.03856}\ } (\bibinfo {year} {2023})}\BibitemShut
  {NoStop}%
\bibitem [{\citenamefont {Slavov}\ and\ \citenamefont
  {Polovnikov}(2023)}]{slavov2023intrachain}%
  \BibitemOpen
  \bibfield  {author} {\bibinfo {author} {\bibfnamefont {B.}~\bibnamefont
  {Slavov}}\ and\ \bibinfo {author} {\bibfnamefont {K.}~\bibnamefont
  {Polovnikov}},\ }\bibfield  {title} {\bibinfo {title} {Intrachain distances
  in a crumpled polymer with random loops},\ }\href@noop {} {\bibfield
  {journal} {\bibinfo  {journal} {JETP Letters}\ }\textbf {\bibinfo {volume}
  {118}},\ \bibinfo {pages} {208} (\bibinfo {year} {2023})}\BibitemShut
  {NoStop}%
\bibitem [{\citenamefont {Cattoni}\ \emph {et~al.}(2017)\citenamefont
  {Cattoni}, \citenamefont {Cardozo~Gizzi}, \citenamefont {Georgieva},
  \citenamefont {Di~Stefano}, \citenamefont {Valeri}, \citenamefont
  {Chamousset}, \citenamefont {Houbron}, \citenamefont {D{\'e}jardin},
  \citenamefont {Fiche}, \citenamefont {Gonz{\'a}lez} \emph
  {et~al.}}]{cattoni2017single}%
  \BibitemOpen
  \bibfield  {author} {\bibinfo {author} {\bibfnamefont {D.~I.}\ \bibnamefont
  {Cattoni}}, \bibinfo {author} {\bibfnamefont {A.~M.}\ \bibnamefont
  {Cardozo~Gizzi}}, \bibinfo {author} {\bibfnamefont {M.}~\bibnamefont
  {Georgieva}}, \bibinfo {author} {\bibfnamefont {M.}~\bibnamefont
  {Di~Stefano}}, \bibinfo {author} {\bibfnamefont {A.}~\bibnamefont {Valeri}},
  \bibinfo {author} {\bibfnamefont {D.}~\bibnamefont {Chamousset}}, \bibinfo
  {author} {\bibfnamefont {C.}~\bibnamefont {Houbron}}, \bibinfo {author}
  {\bibfnamefont {S.}~\bibnamefont {D{\'e}jardin}}, \bibinfo {author}
  {\bibfnamefont {J.-B.}\ \bibnamefont {Fiche}}, \bibinfo {author}
  {\bibfnamefont {I.}~\bibnamefont {Gonz{\'a}lez}}, \emph {et~al.},\ }\bibfield
   {title} {\bibinfo {title} {Single-cell absolute contact probability
  detection reveals chromosomes are organized by multiple low-frequency yet
  specific interactions},\ }\href@noop {} {\bibfield  {journal} {\bibinfo
  {journal} {Nature communications}\ }\textbf {\bibinfo {volume} {8}},\
  \bibinfo {pages} {1} (\bibinfo {year} {2017})}\BibitemShut {NoStop}%
\bibitem [{\citenamefont {Gassler}\ \emph {et~al.}(2017)\citenamefont
  {Gassler}, \citenamefont {Brand{\~a}o}, \citenamefont {Imakaev},
  \citenamefont {Flyamer}, \citenamefont {Ladst{\"a}tter}, \citenamefont
  {Bickmore}, \citenamefont {Peters}, \citenamefont {Mirny},\ and\
  \citenamefont {Tachibana}}]{gassler2017mechanism}%
  \BibitemOpen
  \bibfield  {author} {\bibinfo {author} {\bibfnamefont {J.}~\bibnamefont
  {Gassler}}, \bibinfo {author} {\bibfnamefont {H.~B.}\ \bibnamefont
  {Brand{\~a}o}}, \bibinfo {author} {\bibfnamefont {M.}~\bibnamefont
  {Imakaev}}, \bibinfo {author} {\bibfnamefont {I.~M.}\ \bibnamefont
  {Flyamer}}, \bibinfo {author} {\bibfnamefont {S.}~\bibnamefont
  {Ladst{\"a}tter}}, \bibinfo {author} {\bibfnamefont {W.~A.}\ \bibnamefont
  {Bickmore}}, \bibinfo {author} {\bibfnamefont {J.-M.}\ \bibnamefont
  {Peters}}, \bibinfo {author} {\bibfnamefont {L.~A.}\ \bibnamefont {Mirny}},\
  and\ \bibinfo {author} {\bibfnamefont {K.}~\bibnamefont {Tachibana}},\
  }\bibfield  {title} {\bibinfo {title} {A mechanism of cohesin-dependent loop
  extrusion organizes zygotic genome architecture},\ }\href@noop {} {\bibfield
  {journal} {\bibinfo  {journal} {The EMBO journal}\ }\textbf {\bibinfo
  {volume} {36}},\ \bibinfo {pages} {3600} (\bibinfo {year}
  {2017})}\BibitemShut {NoStop}%
\bibitem [{\citenamefont {Nagano}\ \emph {et~al.}(2013)\citenamefont {Nagano},
  \citenamefont {Lubling}, \citenamefont {Stevens}, \citenamefont
  {Schoenfelder}, \citenamefont {Yaffe}, \citenamefont {Dean}, \citenamefont
  {Laue}, \citenamefont {Tanay},\ and\ \citenamefont
  {Fraser}}]{nagano2013single}%
  \BibitemOpen
  \bibfield  {author} {\bibinfo {author} {\bibfnamefont {T.}~\bibnamefont
  {Nagano}}, \bibinfo {author} {\bibfnamefont {Y.}~\bibnamefont {Lubling}},
  \bibinfo {author} {\bibfnamefont {T.~J.}\ \bibnamefont {Stevens}}, \bibinfo
  {author} {\bibfnamefont {S.}~\bibnamefont {Schoenfelder}}, \bibinfo {author}
  {\bibfnamefont {E.}~\bibnamefont {Yaffe}}, \bibinfo {author} {\bibfnamefont
  {W.}~\bibnamefont {Dean}}, \bibinfo {author} {\bibfnamefont {E.~D.}\
  \bibnamefont {Laue}}, \bibinfo {author} {\bibfnamefont {A.}~\bibnamefont
  {Tanay}},\ and\ \bibinfo {author} {\bibfnamefont {P.}~\bibnamefont
  {Fraser}},\ }\bibfield  {title} {\bibinfo {title} {Single-cell hi-c reveals
  cell-to-cell variability in chromosome structure},\ }\href@noop {} {\bibfield
   {journal} {\bibinfo  {journal} {Nature}\ }\textbf {\bibinfo {volume}
  {502}},\ \bibinfo {pages} {59} (\bibinfo {year} {2013})}\BibitemShut
  {NoStop}%
\bibitem [{\citenamefont {Stevens}\ \emph {et~al.}(2017)\citenamefont
  {Stevens}, \citenamefont {Lando}, \citenamefont {Basu}, \citenamefont
  {Atkinson}, \citenamefont {Cao}, \citenamefont {Lee}, \citenamefont {Leeb},
  \citenamefont {Wohlfahrt}, \citenamefont {Boucher}, \citenamefont
  {O’Shaughnessy-Kirwan} \emph {et~al.}}]{stevens20173d}%
  \BibitemOpen
  \bibfield  {author} {\bibinfo {author} {\bibfnamefont {T.~J.}\ \bibnamefont
  {Stevens}}, \bibinfo {author} {\bibfnamefont {D.}~\bibnamefont {Lando}},
  \bibinfo {author} {\bibfnamefont {S.}~\bibnamefont {Basu}}, \bibinfo {author}
  {\bibfnamefont {L.~P.}\ \bibnamefont {Atkinson}}, \bibinfo {author}
  {\bibfnamefont {Y.}~\bibnamefont {Cao}}, \bibinfo {author} {\bibfnamefont
  {S.~F.}\ \bibnamefont {Lee}}, \bibinfo {author} {\bibfnamefont
  {M.}~\bibnamefont {Leeb}}, \bibinfo {author} {\bibfnamefont {K.~J.}\
  \bibnamefont {Wohlfahrt}}, \bibinfo {author} {\bibfnamefont {W.}~\bibnamefont
  {Boucher}}, \bibinfo {author} {\bibfnamefont {A.}~\bibnamefont
  {O’Shaughnessy-Kirwan}}, \emph {et~al.},\ }\bibfield  {title} {\bibinfo
  {title} {3d structures of individual mammalian genomes studied by single-cell
  hi-c},\ }\href@noop {} {\bibfield  {journal} {\bibinfo  {journal} {Nature}\
  }\textbf {\bibinfo {volume} {544}},\ \bibinfo {pages} {59} (\bibinfo {year}
  {2017})}\BibitemShut {NoStop}%
\bibitem [{\citenamefont {Nagano}\ \emph {et~al.}(2015)\citenamefont {Nagano},
  \citenamefont {Lubling}, \citenamefont {Yaffe}, \citenamefont {Wingett},
  \citenamefont {Dean}, \citenamefont {Tanay},\ and\ \citenamefont
  {Fraser}}]{nagano2015single}%
  \BibitemOpen
  \bibfield  {author} {\bibinfo {author} {\bibfnamefont {T.}~\bibnamefont
  {Nagano}}, \bibinfo {author} {\bibfnamefont {Y.}~\bibnamefont {Lubling}},
  \bibinfo {author} {\bibfnamefont {E.}~\bibnamefont {Yaffe}}, \bibinfo
  {author} {\bibfnamefont {S.~W.}\ \bibnamefont {Wingett}}, \bibinfo {author}
  {\bibfnamefont {W.}~\bibnamefont {Dean}}, \bibinfo {author} {\bibfnamefont
  {A.}~\bibnamefont {Tanay}},\ and\ \bibinfo {author} {\bibfnamefont
  {P.}~\bibnamefont {Fraser}},\ }\bibfield  {title} {\bibinfo {title}
  {Single-cell hi-c for genome-wide detection of chromatin interactions that
  occur simultaneously in a single cell},\ }\href@noop {} {\bibfield  {journal}
  {\bibinfo  {journal} {Nature protocols}\ }\textbf {\bibinfo {volume} {10}},\
  \bibinfo {pages} {1986} (\bibinfo {year} {2015})}\BibitemShut {NoStop}%
\bibitem [{\citenamefont {Ramani}\ \emph {et~al.}(2020)\citenamefont {Ramani},
  \citenamefont {Deng}, \citenamefont {Qiu}, \citenamefont {Lee}, \citenamefont
  {Disteche}, \citenamefont {Noble}, \citenamefont {Shendure},\ and\
  \citenamefont {Duan}}]{ramani2020sci}%
  \BibitemOpen
  \bibfield  {author} {\bibinfo {author} {\bibfnamefont {V.}~\bibnamefont
  {Ramani}}, \bibinfo {author} {\bibfnamefont {X.}~\bibnamefont {Deng}},
  \bibinfo {author} {\bibfnamefont {R.}~\bibnamefont {Qiu}}, \bibinfo {author}
  {\bibfnamefont {C.}~\bibnamefont {Lee}}, \bibinfo {author} {\bibfnamefont
  {C.~M.}\ \bibnamefont {Disteche}}, \bibinfo {author} {\bibfnamefont {W.~S.}\
  \bibnamefont {Noble}}, \bibinfo {author} {\bibfnamefont {J.}~\bibnamefont
  {Shendure}},\ and\ \bibinfo {author} {\bibfnamefont {Z.}~\bibnamefont
  {Duan}},\ }\bibfield  {title} {\bibinfo {title} {Sci-hi-c: a single-cell hi-c
  method for mapping 3d genome organization in large number of single cells},\
  }\href@noop {} {\bibfield  {journal} {\bibinfo  {journal} {Methods}\ }\textbf
  {\bibinfo {volume} {170}},\ \bibinfo {pages} {61} (\bibinfo {year}
  {2020})}\BibitemShut {NoStop}%
\bibitem [{\citenamefont {Galitsyna}\ and\ \citenamefont
  {Gelfand}(2021)}]{galitsyna2021single}%
  \BibitemOpen
  \bibfield  {author} {\bibinfo {author} {\bibfnamefont {A.~A.}\ \bibnamefont
  {Galitsyna}}\ and\ \bibinfo {author} {\bibfnamefont {M.~S.}\ \bibnamefont
  {Gelfand}},\ }\bibfield  {title} {\bibinfo {title} {Single-cell hi-c data
  analysis: safety in numbers},\ }\href@noop {} {\bibfield  {journal} {\bibinfo
   {journal} {Briefings in bioinformatics}\ }\textbf {\bibinfo {volume} {22}},\
  \bibinfo {pages} {bbab316} (\bibinfo {year} {2021})}\BibitemShut {NoStop}%
\bibitem [{\citenamefont {Kos}\ \emph {et~al.}(2021)\citenamefont {Kos},
  \citenamefont {Galitsyna}, \citenamefont {Ulianov}, \citenamefont {Gelfand},
  \citenamefont {Razin},\ and\ \citenamefont
  {Chertovich}}]{kos2021perspectives}%
  \BibitemOpen
  \bibfield  {author} {\bibinfo {author} {\bibfnamefont {P.~I.}\ \bibnamefont
  {Kos}}, \bibinfo {author} {\bibfnamefont {A.~A.}\ \bibnamefont {Galitsyna}},
  \bibinfo {author} {\bibfnamefont {S.~V.}\ \bibnamefont {Ulianov}}, \bibinfo
  {author} {\bibfnamefont {M.~S.}\ \bibnamefont {Gelfand}}, \bibinfo {author}
  {\bibfnamefont {S.~V.}\ \bibnamefont {Razin}},\ and\ \bibinfo {author}
  {\bibfnamefont {A.~V.}\ \bibnamefont {Chertovich}},\ }\bibfield  {title}
  {\bibinfo {title} {Perspectives for the reconstruction of 3d chromatin
  conformation using single cell hi-c data},\ }\href@noop {} {\bibfield
  {journal} {\bibinfo  {journal} {PLoS Computational Biology}\ }\textbf
  {\bibinfo {volume} {17}},\ \bibinfo {pages} {e1009546} (\bibinfo {year}
  {2021})}\BibitemShut {NoStop}%
\bibitem [{\citenamefont {Zhang}\ \emph {et~al.}(2022)\citenamefont {Zhang},
  \citenamefont {Zhou},\ and\ \citenamefont {Ma}}]{zhang2022multiscale}%
  \BibitemOpen
  \bibfield  {author} {\bibinfo {author} {\bibfnamefont {R.}~\bibnamefont
  {Zhang}}, \bibinfo {author} {\bibfnamefont {T.}~\bibnamefont {Zhou}},\ and\
  \bibinfo {author} {\bibfnamefont {J.}~\bibnamefont {Ma}},\ }\bibfield
  {title} {\bibinfo {title} {Multiscale and integrative single-cell hi-c
  analysis with higashi},\ }\href@noop {} {\bibfield  {journal} {\bibinfo
  {journal} {Nature biotechnology}\ }\textbf {\bibinfo {volume} {40}},\
  \bibinfo {pages} {254} (\bibinfo {year} {2022})}\BibitemShut {NoStop}%
\bibitem [{\citenamefont {Ou}\ \emph {et~al.}(2017)\citenamefont {Ou},
  \citenamefont {Phan}, \citenamefont {Deerinck}, \citenamefont {Thor},
  \citenamefont {Ellisman},\ and\ \citenamefont {O’shea}}]{ou2017chromemt}%
  \BibitemOpen
  \bibfield  {author} {\bibinfo {author} {\bibfnamefont {H.~D.}\ \bibnamefont
  {Ou}}, \bibinfo {author} {\bibfnamefont {S.}~\bibnamefont {Phan}}, \bibinfo
  {author} {\bibfnamefont {T.~J.}\ \bibnamefont {Deerinck}}, \bibinfo {author}
  {\bibfnamefont {A.}~\bibnamefont {Thor}}, \bibinfo {author} {\bibfnamefont
  {M.~H.}\ \bibnamefont {Ellisman}},\ and\ \bibinfo {author} {\bibfnamefont
  {C.~C.}\ \bibnamefont {O’shea}},\ }\bibfield  {title} {\bibinfo {title}
  {Chromemt: Visualizing 3d chromatin structure and compaction in interphase
  and mitotic cells},\ }\href@noop {} {\bibfield  {journal} {\bibinfo
  {journal} {Science}\ }\textbf {\bibinfo {volume} {357}},\ \bibinfo {pages}
  {eaag0025} (\bibinfo {year} {2017})}\BibitemShut {NoStop}%
\bibitem [{\citenamefont {Bintu}\ \emph {et~al.}(2018)\citenamefont {Bintu},
  \citenamefont {Mateo}, \citenamefont {Su}, \citenamefont {Sinnott-Armstrong},
  \citenamefont {Parker}, \citenamefont {Kinrot}, \citenamefont {Yamaya},
  \citenamefont {Boettiger},\ and\ \citenamefont {Zhuang}}]{bintu2018super}%
  \BibitemOpen
  \bibfield  {author} {\bibinfo {author} {\bibfnamefont {B.}~\bibnamefont
  {Bintu}}, \bibinfo {author} {\bibfnamefont {L.~J.}\ \bibnamefont {Mateo}},
  \bibinfo {author} {\bibfnamefont {J.-H.}\ \bibnamefont {Su}}, \bibinfo
  {author} {\bibfnamefont {N.~A.}\ \bibnamefont {Sinnott-Armstrong}}, \bibinfo
  {author} {\bibfnamefont {M.}~\bibnamefont {Parker}}, \bibinfo {author}
  {\bibfnamefont {S.}~\bibnamefont {Kinrot}}, \bibinfo {author} {\bibfnamefont
  {K.}~\bibnamefont {Yamaya}}, \bibinfo {author} {\bibfnamefont {A.~N.}\
  \bibnamefont {Boettiger}},\ and\ \bibinfo {author} {\bibfnamefont
  {X.}~\bibnamefont {Zhuang}},\ }\bibfield  {title} {\bibinfo {title}
  {Super-resolution chromatin tracing reveals domains and cooperative
  interactions in single cells},\ }\href@noop {} {\bibfield  {journal}
  {\bibinfo  {journal} {Science}\ }\textbf {\bibinfo {volume} {362}},\ \bibinfo
  {pages} {eaau1783} (\bibinfo {year} {2018})}\BibitemShut {NoStop}%
\bibitem [{\citenamefont {Nir}\ \emph {et~al.}(2018)\citenamefont {Nir},
  \citenamefont {Farabella}, \citenamefont {P{\'e}rez~Estrada}, \citenamefont
  {Ebeling}, \citenamefont {Beliveau}, \citenamefont {Sasaki}, \citenamefont
  {Lee}, \citenamefont {Nguyen}, \citenamefont {McCole}, \citenamefont
  {Chattoraj} \emph {et~al.}}]{nir2018walking}%
  \BibitemOpen
  \bibfield  {author} {\bibinfo {author} {\bibfnamefont {G.}~\bibnamefont
  {Nir}}, \bibinfo {author} {\bibfnamefont {I.}~\bibnamefont {Farabella}},
  \bibinfo {author} {\bibfnamefont {C.}~\bibnamefont {P{\'e}rez~Estrada}},
  \bibinfo {author} {\bibfnamefont {C.~G.}\ \bibnamefont {Ebeling}}, \bibinfo
  {author} {\bibfnamefont {B.~J.}\ \bibnamefont {Beliveau}}, \bibinfo {author}
  {\bibfnamefont {H.~M.}\ \bibnamefont {Sasaki}}, \bibinfo {author}
  {\bibfnamefont {S.~D.}\ \bibnamefont {Lee}}, \bibinfo {author} {\bibfnamefont
  {S.~C.}\ \bibnamefont {Nguyen}}, \bibinfo {author} {\bibfnamefont {R.~B.}\
  \bibnamefont {McCole}}, \bibinfo {author} {\bibfnamefont {S.}~\bibnamefont
  {Chattoraj}}, \emph {et~al.},\ }\bibfield  {title} {\bibinfo {title} {Walking
  along chromosomes with super-resolution imaging, contact maps, and
  integrative modeling},\ }\href@noop {} {\bibfield  {journal} {\bibinfo
  {journal} {PLoS genetics}\ }\textbf {\bibinfo {volume} {14}},\ \bibinfo
  {pages} {e1007872} (\bibinfo {year} {2018})}\BibitemShut {NoStop}%
\bibitem [{\citenamefont {Boettiger}\ and\ \citenamefont
  {Murphy}(2020)}]{boettiger2020advances}%
  \BibitemOpen
  \bibfield  {author} {\bibinfo {author} {\bibfnamefont {A.}~\bibnamefont
  {Boettiger}}\ and\ \bibinfo {author} {\bibfnamefont {S.}~\bibnamefont
  {Murphy}},\ }\bibfield  {title} {\bibinfo {title} {Advances in chromatin
  imaging at kilobase-scale resolution},\ }\href@noop {} {\bibfield  {journal}
  {\bibinfo  {journal} {Trends in Genetics}\ }\textbf {\bibinfo {volume}
  {36}},\ \bibinfo {pages} {273} (\bibinfo {year} {2020})}\BibitemShut
  {NoStop}%
\bibitem [{\citenamefont {Kempfer}\ and\ \citenamefont
  {Pombo}(2020)}]{kempfer2020methods}%
  \BibitemOpen
  \bibfield  {author} {\bibinfo {author} {\bibfnamefont {R.}~\bibnamefont
  {Kempfer}}\ and\ \bibinfo {author} {\bibfnamefont {A.}~\bibnamefont
  {Pombo}},\ }\bibfield  {title} {\bibinfo {title} {Methods for mapping 3d
  chromosome architecture},\ }\href@noop {} {\bibfield  {journal} {\bibinfo
  {journal} {Nature Reviews Genetics}\ }\textbf {\bibinfo {volume} {21}},\
  \bibinfo {pages} {207} (\bibinfo {year} {2020})}\BibitemShut {NoStop}%
\bibitem [{\citenamefont {Su}\ \emph {et~al.}(2020)\citenamefont {Su},
  \citenamefont {Zheng}, \citenamefont {Kinrot}, \citenamefont {Bintu},\ and\
  \citenamefont {Zhuang}}]{su2020genome}%
  \BibitemOpen
  \bibfield  {author} {\bibinfo {author} {\bibfnamefont {J.-H.}\ \bibnamefont
  {Su}}, \bibinfo {author} {\bibfnamefont {P.}~\bibnamefont {Zheng}}, \bibinfo
  {author} {\bibfnamefont {S.~S.}\ \bibnamefont {Kinrot}}, \bibinfo {author}
  {\bibfnamefont {B.}~\bibnamefont {Bintu}},\ and\ \bibinfo {author}
  {\bibfnamefont {X.}~\bibnamefont {Zhuang}},\ }\bibfield  {title} {\bibinfo
  {title} {Genome-scale imaging of the 3d organization and transcriptional
  activity of chromatin},\ }\href@noop {} {\bibfield  {journal} {\bibinfo
  {journal} {Cell}\ }\textbf {\bibinfo {volume} {182}},\ \bibinfo {pages}
  {1641} (\bibinfo {year} {2020})}\BibitemShut {NoStop}%
\bibitem [{\citenamefont {Liu}\ \emph {et~al.}(2020)\citenamefont {Liu},
  \citenamefont {Lu}, \citenamefont {Yang}, \citenamefont {Chen}, \citenamefont
  {Radda}, \citenamefont {Hu}, \citenamefont {Katz},\ and\ \citenamefont
  {Wang}}]{liu2020multiplexed}%
  \BibitemOpen
  \bibfield  {author} {\bibinfo {author} {\bibfnamefont {M.}~\bibnamefont
  {Liu}}, \bibinfo {author} {\bibfnamefont {Y.}~\bibnamefont {Lu}}, \bibinfo
  {author} {\bibfnamefont {B.}~\bibnamefont {Yang}}, \bibinfo {author}
  {\bibfnamefont {Y.}~\bibnamefont {Chen}}, \bibinfo {author} {\bibfnamefont
  {J.~S.}\ \bibnamefont {Radda}}, \bibinfo {author} {\bibfnamefont
  {M.}~\bibnamefont {Hu}}, \bibinfo {author} {\bibfnamefont {S.~G.}\
  \bibnamefont {Katz}},\ and\ \bibinfo {author} {\bibfnamefont
  {S.}~\bibnamefont {Wang}},\ }\bibfield  {title} {\bibinfo {title}
  {Multiplexed imaging of nucleome architectures in single cells of mammalian
  tissue},\ }\href@noop {} {\bibfield  {journal} {\bibinfo  {journal} {Nature
  communications}\ }\textbf {\bibinfo {volume} {11}},\ \bibinfo {pages} {1}
  (\bibinfo {year} {2020})}\BibitemShut {NoStop}%
\bibitem [{\citenamefont {Xie}\ and\ \citenamefont
  {Liu}(2021)}]{xie2021single}%
  \BibitemOpen
  \bibfield  {author} {\bibinfo {author} {\bibfnamefont {L.}~\bibnamefont
  {Xie}}\ and\ \bibinfo {author} {\bibfnamefont {Z.}~\bibnamefont {Liu}},\
  }\bibfield  {title} {\bibinfo {title} {Single-cell imaging of genome
  organization and dynamics},\ }\href@noop {} {\bibfield  {journal} {\bibinfo
  {journal} {Molecular Systems Biology}\ }\textbf {\bibinfo {volume} {17}},\
  \bibinfo {pages} {e9653} (\bibinfo {year} {2021})}\BibitemShut {NoStop}%
\bibitem [{\citenamefont {Li}\ \emph {et~al.}(2021)\citenamefont {Li},
  \citenamefont {Eshein}, \citenamefont {Virk}, \citenamefont {Eid},
  \citenamefont {Wu}, \citenamefont {Frederick}, \citenamefont {VanDerway},
  \citenamefont {Gladstein}, \citenamefont {Huang}, \citenamefont {Shim} \emph
  {et~al.}}]{li2021nanoscale}%
  \BibitemOpen
  \bibfield  {author} {\bibinfo {author} {\bibfnamefont {Y.}~\bibnamefont
  {Li}}, \bibinfo {author} {\bibfnamefont {A.}~\bibnamefont {Eshein}}, \bibinfo
  {author} {\bibfnamefont {R.~K.}\ \bibnamefont {Virk}}, \bibinfo {author}
  {\bibfnamefont {A.}~\bibnamefont {Eid}}, \bibinfo {author} {\bibfnamefont
  {W.}~\bibnamefont {Wu}}, \bibinfo {author} {\bibfnamefont {J.}~\bibnamefont
  {Frederick}}, \bibinfo {author} {\bibfnamefont {D.}~\bibnamefont
  {VanDerway}}, \bibinfo {author} {\bibfnamefont {S.}~\bibnamefont
  {Gladstein}}, \bibinfo {author} {\bibfnamefont {K.}~\bibnamefont {Huang}},
  \bibinfo {author} {\bibfnamefont {A.~R.}\ \bibnamefont {Shim}}, \emph
  {et~al.},\ }\bibfield  {title} {\bibinfo {title} {Nanoscale chromatin imaging
  and analysis platform bridges 4d chromatin organization with molecular
  function},\ }\href@noop {} {\bibfield  {journal} {\bibinfo  {journal}
  {Science advances}\ }\textbf {\bibinfo {volume} {7}},\ \bibinfo {pages}
  {eabe4310} (\bibinfo {year} {2021})}\BibitemShut {NoStop}%
\bibitem [{\citenamefont {Gabriele}\ \emph {et~al.}(2022)\citenamefont
  {Gabriele}, \citenamefont {Brand{\~a}o}, \citenamefont {Grosse-Holz},
  \citenamefont {Jha}, \citenamefont {Dailey}, \citenamefont {Cattoglio},
  \citenamefont {Hsieh}, \citenamefont {Mirny}, \citenamefont {Zechner},\ and\
  \citenamefont {Hansen}}]{gabriele2022dynamics}%
  \BibitemOpen
  \bibfield  {author} {\bibinfo {author} {\bibfnamefont {M.}~\bibnamefont
  {Gabriele}}, \bibinfo {author} {\bibfnamefont {H.~B.}\ \bibnamefont
  {Brand{\~a}o}}, \bibinfo {author} {\bibfnamefont {S.}~\bibnamefont
  {Grosse-Holz}}, \bibinfo {author} {\bibfnamefont {A.}~\bibnamefont {Jha}},
  \bibinfo {author} {\bibfnamefont {G.~M.}\ \bibnamefont {Dailey}}, \bibinfo
  {author} {\bibfnamefont {C.}~\bibnamefont {Cattoglio}}, \bibinfo {author}
  {\bibfnamefont {T.-H.~S.}\ \bibnamefont {Hsieh}}, \bibinfo {author}
  {\bibfnamefont {L.}~\bibnamefont {Mirny}}, \bibinfo {author} {\bibfnamefont
  {C.}~\bibnamefont {Zechner}},\ and\ \bibinfo {author} {\bibfnamefont {A.~S.}\
  \bibnamefont {Hansen}},\ }\bibfield  {title} {\bibinfo {title} {Dynamics of
  ctcf-and cohesin-mediated chromatin looping revealed by live-cell imaging},\
  }\href@noop {} {\bibfield  {journal} {\bibinfo  {journal} {Science}\ }\textbf
  {\bibinfo {volume} {376}},\ \bibinfo {pages} {496} (\bibinfo {year}
  {2022})}\BibitemShut {NoStop}%
\bibitem [{\citenamefont {Fudenberg}\ \emph {et~al.}(2017)\citenamefont
  {Fudenberg}, \citenamefont {Abdennur}, \citenamefont {Imakaev}, \citenamefont
  {Goloborodko},\ and\ \citenamefont {Mirny}}]{fudenberg2017emerging}%
  \BibitemOpen
  \bibfield  {author} {\bibinfo {author} {\bibfnamefont {G.}~\bibnamefont
  {Fudenberg}}, \bibinfo {author} {\bibfnamefont {N.}~\bibnamefont {Abdennur}},
  \bibinfo {author} {\bibfnamefont {M.}~\bibnamefont {Imakaev}}, \bibinfo
  {author} {\bibfnamefont {A.}~\bibnamefont {Goloborodko}},\ and\ \bibinfo
  {author} {\bibfnamefont {L.~A.}\ \bibnamefont {Mirny}},\ }\bibfield  {title}
  {\bibinfo {title} {Emerging evidence of chromosome folding by loop
  extrusion},\ }in\ \href@noop {} {\emph {\bibinfo {booktitle} {Cold Spring
  Harbor symposia on quantitative biology}}},\ Vol.~\bibinfo {volume} {82}\
  (\bibinfo {organization} {Cold Spring Harbor Laboratory Press},\ \bibinfo
  {year} {2017})\ pp.\ \bibinfo {pages} {45--55}\BibitemShut {NoStop}%
\bibitem [{\citenamefont {Goloborodko}\ \emph {et~al.}(2016)\citenamefont
  {Goloborodko}, \citenamefont {Marko},\ and\ \citenamefont
  {Mirny}}]{Goloborodko_2016}%
  \BibitemOpen
  \bibfield  {author} {\bibinfo {author} {\bibfnamefont {A.}~\bibnamefont
  {Goloborodko}}, \bibinfo {author} {\bibfnamefont {J.~F.}\ \bibnamefont
  {Marko}},\ and\ \bibinfo {author} {\bibfnamefont {L.~A.}\ \bibnamefont
  {Mirny}},\ }\bibfield  {title} {\bibinfo {title} {Chromosome compaction by
  active loop extrusion},\ }\href@noop {} {\bibfield  {journal} {\bibinfo
  {journal} {Biophysical Journal}\ }\textbf {\bibinfo {volume} {110}},\
  \bibinfo {pages} {2162} (\bibinfo {year} {2016})}\BibitemShut {NoStop}%
\bibitem [{\citenamefont {Grosberg}\ and\ \citenamefont
  {Khokhlov}(1994)}]{GKh_1994}%
  \BibitemOpen
  \bibfield  {author} {\bibinfo {author} {\bibfnamefont {A.~Y.}\ \bibnamefont
  {Grosberg}}\ and\ \bibinfo {author} {\bibfnamefont {A.}~\bibnamefont
  {Khokhlov}},\ }\href@noop {} {\emph {\bibinfo {title} {Statistical Physics of
  Macromolecules}}}\ (\bibinfo  {publisher} {Woodbury, NY: AIP Press},\
  \bibinfo {year} {1994})\BibitemShut {NoStop}%
\bibitem [{\citenamefont {Zhang}\ and\ \citenamefont
  {Xia}(2001)}]{zhang2001probability}%
  \BibitemOpen
  \bibfield  {author} {\bibinfo {author} {\bibfnamefont {L.-X.}\ \bibnamefont
  {Zhang}}\ and\ \bibinfo {author} {\bibfnamefont {A.}~\bibnamefont {Xia}},\
  }\bibfield  {title} {\bibinfo {title} {Probability of triple contacts in
  polymer chains},\ }\href@noop {} {\bibfield  {journal} {\bibinfo  {journal}
  {European polymer journal}\ }\textbf {\bibinfo {volume} {37}},\ \bibinfo
  {pages} {1263} (\bibinfo {year} {2001})}\BibitemShut {NoStop}%
\bibitem [{\citenamefont {Nuebler}\ \emph {et~al.}(2018)\citenamefont
  {Nuebler}, \citenamefont {Fudenberg}, \citenamefont {Imakaev}, \citenamefont
  {Abdennur},\ and\ \citenamefont {Mirny}}]{nuebler2018chromatin}%
  \BibitemOpen
  \bibfield  {author} {\bibinfo {author} {\bibfnamefont {J.}~\bibnamefont
  {Nuebler}}, \bibinfo {author} {\bibfnamefont {G.}~\bibnamefont {Fudenberg}},
  \bibinfo {author} {\bibfnamefont {M.}~\bibnamefont {Imakaev}}, \bibinfo
  {author} {\bibfnamefont {N.}~\bibnamefont {Abdennur}},\ and\ \bibinfo
  {author} {\bibfnamefont {L.}~\bibnamefont {Mirny}},\ }\bibfield  {title}
  {\bibinfo {title} {Chromatin organization by an interplay of loop extrusion
  and compartmental segregation},\ }\href@noop {} {\bibfield  {journal}
  {\bibinfo  {journal} {Biophysical Journal}\ }\textbf {\bibinfo {volume}
  {114}},\ \bibinfo {pages} {30a} (\bibinfo {year} {2018})}\BibitemShut
  {NoStop}%
\bibitem [{\citenamefont {Hafner}\ and\ \citenamefont
  {Boettiger}(2022)}]{hafner2022spatial}%
  \BibitemOpen
  \bibfield  {author} {\bibinfo {author} {\bibfnamefont {A.}~\bibnamefont
  {Hafner}}\ and\ \bibinfo {author} {\bibfnamefont {A.}~\bibnamefont
  {Boettiger}},\ }\bibfield  {title} {\bibinfo {title} {The spatial
  organization of transcriptional control},\ }\href@noop {} {\bibfield
  {journal} {\bibinfo  {journal} {Nature Reviews Genetics}\ ,\ \bibinfo {pages}
  {1}} (\bibinfo {year} {2022})}\BibitemShut {NoStop}%
\end{thebibliography}%

\end{document}